\pdfoutput=1
\documentclass[useAMS,usenatbib]{mn2e}
\usepackage{amsmath}
\usepackage{amssymb}

\def\etal{{et al.}}
\def\gtwodeff{g_{2D\!,e\!\;\!f\!f}}

\def\uK{\mu\mbox{K}}
\def\deg{^\circ}
\def\Bmode{{\it B}-mode}
\def\Bmodes{{\it B}-modes}
\def\Emode{{\it E}-mode}
\def\Emodes{{\it E}-modes}

\usepackage{graphicx}

\title[Genus Topology and Cross-Correlation of BICEP2 with Planck 353
  GHz \Bmodes:  Further Evidence for Gravity Wave Detection]{Genus
  Topology and Cross-Correlation of BICEP2 and Planck 353 GHz B-Modes:  Further Evidence Favoring Gravity Wave Detection}
\author[Wesley N. Colley and J. Richard Gott, III]{Wesley N. Colley$^{1}$\thanks{E-mail:
colleyw@uah.edu (WNC); jrg@astro.princeton.edu (JRG)} and J. Richard
Gott, III$^{2}$\footnotemark[1]\\
$^{1}$University of Alabama in Huntsville Center for Modeling, Simulation and Analysis, Huntsville, AL 35899, USA\\
$^{2}$Princeton University Department of Astrophysics Sciences, Princeton, NJ 08544, USA} 
\begin{document}

\date{Accepted 2014 December 01. Received 2014 November 25; in original form 2014 September 14}

\pagerange{\pageref{firstpage}--\pageref{lastpage}} \pubyear{????}

\maketitle

\label{firstpage}

\begin{abstract}

We have analyzed the genus topology of the BICEP2 \Bmodes\ and find
them to be Gaussian random phase as expected if they have a
cosmological origin.  These BICEP2 \Bmodes\ can be produced by gravity
waves in the early universe, but question has arisen as to
whether these \Bmodes\ (for $50 < l < 120$) may instead be produced by
foreground polarized dust emission.  The dust emission at 150 GHz
observed by BICEP2 should be less in magnitude but similar in
structure to that at 353 GHz.  We have therefore calculated and mapped
the \Bmodes\ in the BICEP2 region from the publicly available {\it Q}
and {\it U} 353 GHz preliminary Planck polarization maps.  These have
a genus curve that is different from that seen in the BICEP2
observations, with features at different locations from those in the
BICEP2 map.  The two maps show a positive correlation coefficient of
15.2\% $\pm$ 3.9\% ($1\sigma$).  This requires the amplitude of the
Planck ($50 < l <120$) dust modes to be low in the BICEP2 region, and
the majority of the Planck 353 GHz signal in the BICEP2 region in
these modes to be noise.  We can explain the observed correlation
coefficient of 15.2\% with a BICEP2 gravity wave signal with an rms
amplitude equal to 54\% of the total BICEP2 rms amplitude.  The
gravity wave signal corresponds to a tensor-to-scalar ratio $r = 0.11
\pm 0.04$ ($1\sigma$).  This is consistent with a gravity wave signal
having been detected at a $2.5\sigma$ level.  The Planck and BICEP2
teams have recently engaged in joint analysis of their combined
data---it will be interesting to see if that collaboration reaches
similar conclusions.

\end{abstract}

\begin{keywords}

cosmology: cosmic background radiation---cosmology:
observations---cosmology: cosmological
parameters---methods: statistical

\end{keywords}

\section{Introduction}

The BICEP2 team has announced discovery of {\it B} polarization modes
on angular scales of $1\deg$ to $5\deg$ ($l = 40$ to $l = 200$) which
are of an amplitude and angular scale that are too large to be due to
gravitational lensing (BICEP2 Collaboration 2014).  These \Bmodes\ are
just in line with what would be expected from gravity waves from the
early universe (Seljak \& Zaldarriaga 1997, Kamionkowski \etal\ 1997).
Their results suggest that if dust contamination is small (they have
picked a region of the sky with very little foreground dust), the
tensor-to-scalar ratio is $r = 0.2$ as compared with a value of $r =
0.13$ expected from simple single-field slow-roll chaotic inflation
(Linde 1983) with a simple quadratic potential $V(\phi) =
(1/2)m^2\phi^2$ representing a simple massive scalar field (with mass
$m$)---for possible implications for dark energy, see Slepian~\etal\
2014, Gott \& Slepian 2011.  The BICEP2
\Bmode\ map is several $\sigma$ in terms of signal-to-noise, and they
find the same modes however they rotate their telescope and whether
they consider the first or last half of their data.  There thus seems
little question that this signal is on the sky and not an instrumental
effect.  A simulation they present shows that the \Bmodes\ observed
have significantly larger amplitude than would be expected from
gravitational lensing.  The main question seems to be whether the
\Bmodes\ could instead be due entirely to \Bmodes\ produced by
foreground dust.  The BICEP2 team estimates that the dust
contamination is low corresponding to a false value of $r = 0.04$ at
most, and that their signal is produced by gravity waves with $r >
0.16$.  The BICEP2 region was chosen to be one of the least
contaminated by dust in the entire sky.  Mortonson, and Seljak (2014)
and Flauger, Hill \& Spergel (2014) argue that given the uncertainties
of the amplitude of the dust polarization at the BICEP2 frequency of
150 GHz one cannot say conclusively at present whether the
\Bmodes\ detected by BICEP2 are due to gravity waves or just polarized
dust.  All of these studies looked at the power spectrum of the
\Bmodes.  Flauger, Hill \& Spergel (2014) in particular, fitting the
\Bmodes\ power spectrum on the $1\deg$ to $5\deg$ scale, find that a
model with $r = 0.2$ and no appreciable dust polarization (reduced
$\chi^2 = 1.1$) is acceptable, as well as a model with $r = 0$ and
dust \Bmodes\ (reduced $\chi^2 = 1.7$).  They thus conclude that given
the present uncertainty in the amplitude of the dust emission
\Bmodes\ at 150 GHz one cannot say at present whether the BICEP2
\Bmodes\ are due to gravity waves or dust polarization.  Those authors
have digitized a publicly available Planck polarization map to compare
with the BICEP2 map.  We will similarly digitize and utilize this
publicly available Planck polarization map in our study.

\section[]{Study Design}
The main uncertainty seems to be the amplitude of the dust signal in
the BICEP2 map.  Therefore we designed a study to test between a
gravity wave versus dust origin for the BICEP2 \Bmodes\ that does not
depend on the amplitude of the dust signal at all.

We have previously used genus statistics to study the 3D topology of
large scale structure and the 2D topology of hot and cold spots in the
microwave background.  The results of all these studies have been
consistent with Gaussian random phase initial conditions demanded by
inflation (Gott~\etal\ 2007, Colley~\etal\ 2003, Colley~\etal\ 1996).
In the cosmic microwave background our genus statistic is $g(\nu)$ =
number of hot spots $-$ number of cold spots, where $\nu$ is the number
of standard deviations above the mean temperature.  For a Gaussian
random phase distribution $g(\nu) = \nu\exp(-\nu^2/2)$.

We (Colley \& Gott 2003) have found that the genus topology of the
WMAP temperature field is in agreement with the Gaussian random phase
model.  Departures from a Gaussian random phase distribution can be
quantified by the parameter $f_{NL}$ invented by Komatsu \& Spergel
(2000).  A perfect Gaussian random phase distribution would have
$f_{NL} = 0$.  The smallest detectable levels from the CMB would be of
order 5.  Standard slow-roll inflation (a simple field rolling down a
hill) predicts values of $f_{NL}$ of $10^{-2}$ to $10^{-1}$ (close to
zero and undetectable with current data) according to calculations by
Maldacena (2005) and others.  For gravity waves Maldacena and Pimentel
(2011) conclude that $f_{NL} \sim 1$, again essentially undetectable.
Thus, standard inflation predicts values of $f_{NL}$ near
zero--consistent with a Gaussian random phase distribution.  For
comparison, the COBE results set 68\% confidence limits of $-1500 <
f_{NL} < 1500$.  Using 2D genus topology on the WMAP data, we were
able to improve these limits to $-101 < f_{NL} < 107$ at the 95\%
confidence level (Gott~\etal\ 2007).  The WMAP team found $-58 <
f_{NL} < 134$ with an independent analysis (Spergel~\etal\ 2007).  All
these ranges are consistent with $f_{NL}$ near zero.  The Planck
satellite, drawing upon a much higher resolution map, and testing for
random phases with a different (bi-spectrum) method, has recently
found $-3.1 < f_{NL} < 8.5$ (at 68\% confidence) (Planck Collaboration
2013), again consistent with $f_{NL} = 0$ and the predictions of
standard inflation.  Gravity waves should thus produce Gaussian random
phase fields to our limits of detection with 2D genus topology.

The differential operators used to calculate the \Bmodes\, when
applied to a Gaussian random field, yield a Gaussian random field.
This is particularly clear when one considers this in spherical
harmonics where the operators applied to the {\it Q} and
{\it U} maps are calculated by multiplying $a_{lm}$'s by factors
involving $l$, but the Gaussian form of their amplitude distributions
and their random phases are not altered.  Therefore, if the
\Bmodes\ from BICEP2 are cosmological, due to gravity waves in the
early universe we expect their genus curves to be consistent with the
Gaussian random phase formula.  If that is observed it will favor
gravity waves over dust polarization because dust polarization in a
low dust region is not necessarily guaranteed to be Gaussian random
phase, consisting of perhaps only a few polarizing sheets of dust in
the line of sight.  This phase of the study uses only the BICEP2 data
with no reliance on Planck data at all.

In the second phase of the study, we will use the publicly available
Planck data at 353 GHz (Stokes polarization parameters {\it Q} and
{\it U}) to compute the {\it B} polarization modes.  At this frequency,
polarized \Bmode\ emission is surely dominated by dust; so we will
assume that any signal in the {\it B} polarization modes is due to
dust.  We will then compare the genus curve for the Planck 353 GHz map
to that seen in the BICEP2 data.  If they agree, that is evidence in
favor of $r = 0$ and dust polarization only.  In addition we will
compare the 353 GHz \Bmode\ map with the BICEP2 \Bmode\ map.  If the
two agree with positive and negative (clockwise and counterclockwise
swirls in polarization) regions at the same locations this would
constitute a proof that the BICEP2 \Bmode\ polarization was due to
dust and not cosmological.  It would falsify the claim that the
particular \Bmodes\ seen in the BICEP2 map were due to gravity waves.
We are making no specific assumption about the amplitude of the
polarization at 150 GHz, just that the dust is in the same locations
and that the polarization angles are similar at the two frequencies.
If the \Bmodes\ from the dust are non-Gaussian random phase, all the
better.  We expect the \Bmodes\ from the dust to produce a complicated
map, which we will compare directly to the \Bmodes\ in the BICEP2 data
to look for coincidences.  If all the features detected in the BICEP2
\Bmode\ map are explained by features already found in the Planck dust
dominated \Bmode\ map the detection of gravity waves will be
falsified.

The study is designed to go either way: supporting the detection of
gravity waves if the BICEP2 \Bmode\ map is Gaussian random phase and
quite different from the Planck 353 GHz \Bmode\ map, and refuting the
detection if the BICEP2 and Planck 353 GHz \Bmode\ maps are nearly identical.

Previous studies have considered the power spectrum of the \Bmodes.
But this leaves out the other information in the maps.  One of us
(WNC), in his thesis produced two maps with identical power spectra,
where one was Gaussian random phase and the other showed a Taco Bell
logo (Colley 1998).  The only difference was the non-random phase
nature of the second picture.  The first picture looked like noise,
and in the second one could see a blurry Taco Bell sign.  The genus
statistic easily distinguished between these two; the first fit the
Gaussian random phase formula $g(\nu) = \nu\exp(-\nu^2/2)$, while the
second just counted the particular cold spots for the letters and
stylized bell in the logo over a wide range of $\nu$, and was not in
agreement with the random phase formula.  Looking at the power
spectrum alone is insufficient.

It is important whether the \Bmodes\ detected by BICEP2
show up in the Planck dust polarization \Bmode\ maps or not.  Note we
are only considering the locations of the features not their
amplitude.  And we are only making use of the publicly available
BICEP2 data and Planck data that were already utilized by the BICEP2
team and Flauger, Hill, and Spergel (2014).  We are just using them in
a different, and complimentary way.

\section[]{Using the BICEP2 Maps}

To create a usable version of the BICEP2 data, we first had to convert
the figures from the PDF document (BICEP2 Collaboration 2014b) into
usable data, showing the amplitude of the \Bmodes\ as a function of
position on the sky.  The PDF shows the {\it E-} and \Bmode\ signals
observed by BICEP and simulated {\it E} and {\it B} maps of expected
backgrounds, chiefly gravitational lensing.  First, we simply
displayed the PDF such that each figure was at nearly the full
resolution of the screen (a standard 1920$\times$1080 HD display), and
used a screen grabber to generate a PNG image.  These maps present the
major difficulty in that they are shown with black ``headless'' vector
indicators of the polarization direction.  To handle this, we created
a custom Java program that first displays the trace of the color table
(from the scale bar) on top of three two-dimensional histograms of
pixel values in the map, where the axes of these histograms are the R
and G levels, the G and B levels and the B and R levels in the RGB
pixel data.  What was quite apparent was that the majority of the
pixel data indeed lay along the trace of the color table from the
scale bar, but that there were also ``echoes'' of this trace at lower
intensity for a much lower number of pixels.  These turn out to be the
pixels where there is anti-aliasing of the black headless vector
segments.  Fortunately, these echoes are sufficiently separated from
the full intensity color table trace that one can readily identify
which pixels have the full intensity, and which do not (in practice,
for each pixel we found the minimum RGB distance from segments along
the natural color table trace, and any with a minimum distance of
greater than 4 was regarded to be corrupted by headless vector segment
or its anti-aliasing pixels).  For each excluded pixel, we searched to
find the nearest 3 non-excluded pixels that geometrically enclosed the
original pixel.  This forms three triangles in the spaces
$(x,y,\mbox{R})$, $(x,y,\mbox{G})$ and $(x,y,\mbox{B})$.  Treating
each triangle as a plane, we interpolated to the original pixel
location and used that interpolated value to replace the original
excluded pixel value.  The natural outputs are PNG maps on which the
headless vectors have been removed.  However, because we have
referenced to the the scale bar color table in this process, we can
also output each map's values in physical units.

We now have physical-unit maps for each of the four original map
images ({\it E}, {\it B}, simulated {\it E} and simulated {\it B}).
For analysis, we smooth these maps with a Gaussian kernel with $\sigma
= 5$ pixels.  This not only smooths over the blocky appearance of the
original BICEP2 maps, but also smooths over the generally minor
artifacts arising from the headless vector removal process.

\begin{figure}
\includegraphics[width=3.25in]{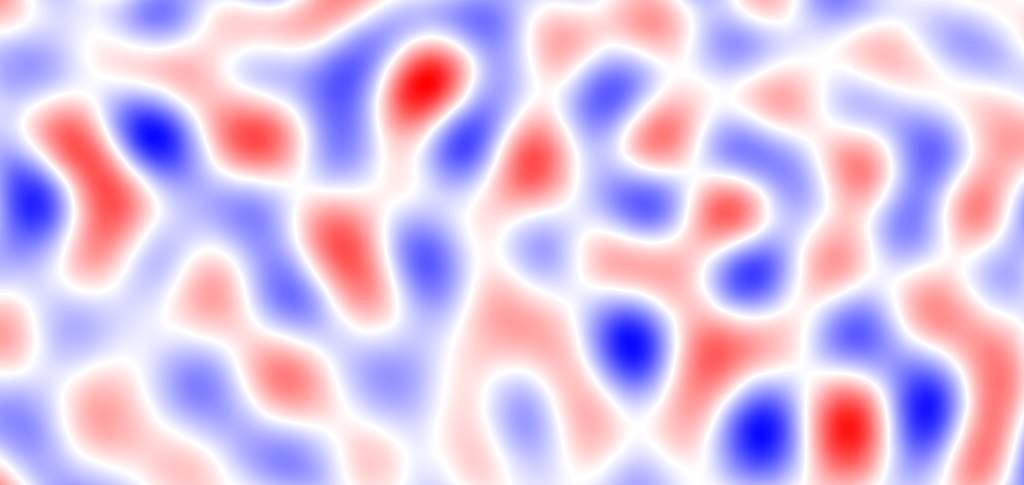}
 \caption{BICEP2 \Bmodes\ in Mercator projection in the region
   $|\mbox{RA}| \leq 30^\circ$, $-65^\circ \leq \mbox{Dec} \leq
   -50^\circ$.  The stretch is from $-0.271\uK$ (blue) to $0.271\uK$ (red).}
\label{bicepBMerc}
\end{figure}

The next problem is that of the sensitivity, which is non-uniform
across the maps.  Fortunately, the BICEP2 Collaboration (2014c) has
provided a map over their sensitivity in Fig.~24.  We used this
sensitivity map to normalize the BICEP \Bmode\ data, and cut our
analysis region to $|\mbox{RA}| \leq 30^\circ$, $-65^\circ \leq
\mbox{Dec} \leq -50^\circ$, inside of which the lowest weight was 
roughly 0.3.  Finally, we projected this flattened map into Mercator
(Fig.~\ref{bicepBMerc}) and Lambert equal-area cylindrical
projections.  Using two different projections was necessary---for
statistical measurements on the map, we chose the equal-area
projection, but for the genus, we chose the Mercator projection.
The conformal Mercator projection preserves angles on the sphere
locally.  This ensures that the density contour surfaces meet the
survey boundary at the correct angles.  The color scheme in
Fig.~\ref{bicepBMerc} is one we used in Gott~\etal\ (2007).  White is
a \Bmode\ of zero.  Red ink indicates positive \Bmode\ with the amount
of red ink per pixel proportional to the value of the positive \Bmode\
at that location.  Blue ink indicates negative \Bmode\ with the amount
of blue ink per pixel proportional to the amount of negative \Bmode\ at
that location.

\section[]{Development of the Dust \Bmode\ Map}

The Planck team (Boulanger 2014) has provided the {\it I}, {\it Q} and
{\it U} maps of dust polarzation.  The {\it Q} and {\it U} maps are
presented in a Mollweide projection, with color table scales given
below each.  The first task is, therefore, to convert the color table
images back to physical units.  Our process is quite similar to that
described for the BICEP2 map above, except without the pesky headless
vectors to clean up.  We first displayed the PDF and did a simple
screen grab at a high zoom level, to convert the PDF data to standard
PNG pixel data.  In IDL, at each pixel location, we found in RGB
vector space the three RGB values from the color scale that minimized
the distance between themselves and the pixel's RGB values.  We then fit a
quadratic function to the squared RGB distances.  Minimizing the
quadratic gave us an interpolated locus on the color scale, which we
translated to temperature using the scale's stated graduation.  The
result was a Mollweide projection of both the {\it Q} and {\it U}
polarization maps in physical units, namely Kelvins.  But to compare
with the BICEP2 \Bmode\ map, we obviously needed to convert the {\it
Q} and {\it U} to {\it B}.

For the \Bmode\ analysis, we used the HEALPix library (e.g.,
G\'orski~\etal\ 2005).  Using the IDL interface, we iterated over all
the HEALPix pixels at a resolution level of $n_{side} = 512$, and
projected them back to the Mollweide map to find the location in our {\it Q}
and {\it U} Mollweide maps (we simply used bilinear interpolation to
estimate the temperature value at a particular HEALPix pixel
location).  We then had a HEALPix map for the {\it Q} and {\it U} dust maps.

\begin{figure}
\includegraphics[width=3.25in]{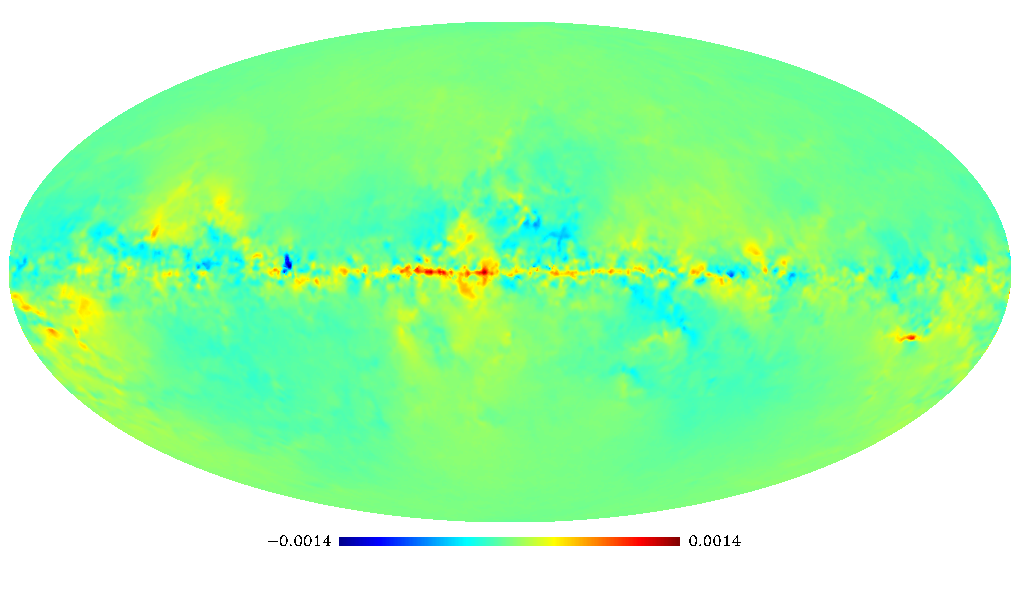}
 \caption{
All-sky map of the \Bmodes\ from the publicly available Planck {\it Q} and
{\it U} maps we have digitized and where we have computed the {\it B}
polarization modes.  The units in the scale
measure fluctuations in brightness temperature (K). 
}
\label{dustB}
\end{figure}

Computing the \Bmode\ values was straight-forward with the standard
modules inside the HEALPix library; the internal mathematical
techniques are described by Guzik~\etal\ (2000).  Our approach was to
develop a C++ program which loaded the HEALPix {\it Q} and {\it U}
maps, and then called the {\tt map2alm\_spin} routine, whose
documentation (Hivon 2010) explains that by calling the routine with a
spin argument of 2, it takes as input the {\it Q} and {\it U} maps,
and outputs the spin zero $a_{lm}$ values for the {\it E-} and
\Bmodes\ associated with the {\it Q} and {\it U} polarizations.  From
there, it is just a matter of using the usual (spin zero) {\tt
  alm2map} routine to convert $a_{lm}$ values into a sky map of the
\Bmodes\ of the dust, as observed by Planck.  Fig.~\ref{dustB} shows
the result of this process.  Fig.~\ref{dustBMerc} shows the dust map
in just the BICEP2 region (as a Mercator projection).

The Planck \Bmode\ map described above (Fig.~\ref{dustBMerc}) clearly looks
nothing like the BICEP2 \Bmode\ map.  However, the BICEP2 team
conducted some data processing that one must mimic in order to conduct
statistical analyses.  In particular, the BICEP2 map only shows modes
from $50 < l < 120$ and has been ``desplined'' as we shall describe.  We
will use 5 somewhat different techniques with various degrees of
complexity for achieving this.

\begin{figure}
\resizebox{\hsize}{!}{\includegraphics{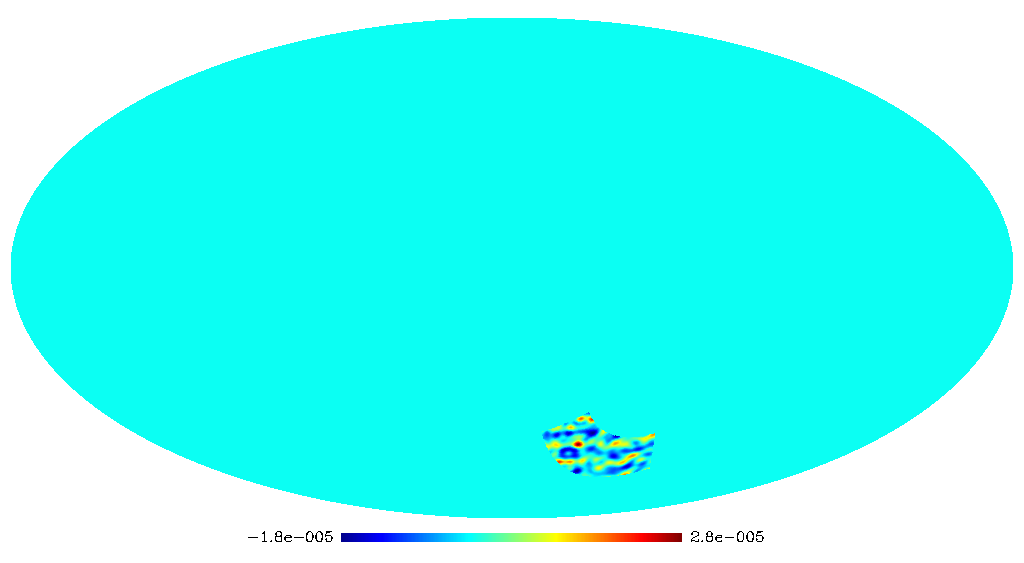}}
 \caption{Planck 353 GHz \Bmodes\ in the BICEP2 analysis region, using $50
   < l < 120$ {\it B} (in K).  Note the striations.}
\label{dustTruncBicepReg}
\end{figure}

\begin{figure}
\resizebox{\hsize}{!}{\includegraphics{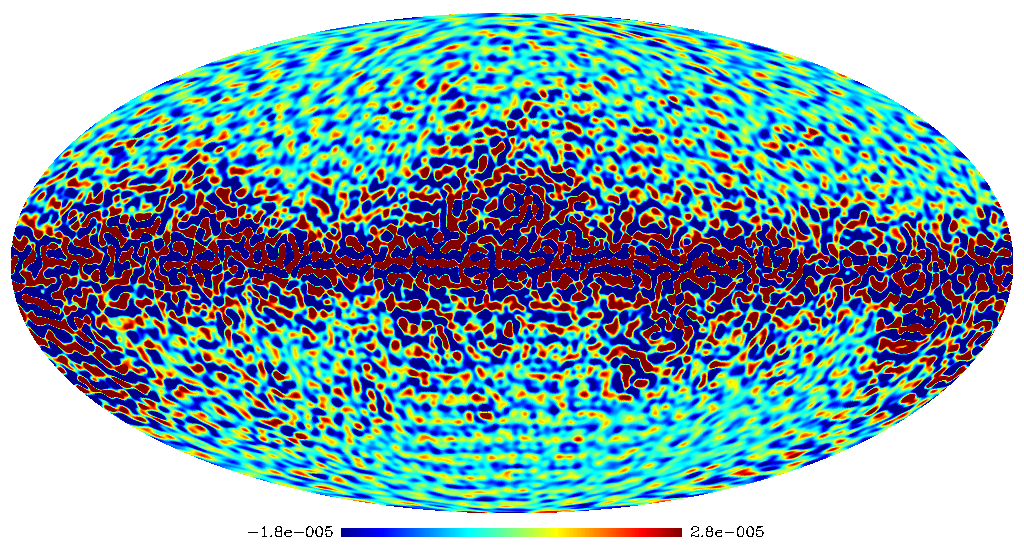}}
 \caption{Planck 353 GHz \Bmodes\ for the full sky, using $50
   < l < 120$ {\it B} (in K), stretched as
   in Fig.~\ref{dustTruncBicepReg}.   This stretch saturates high values near
   the plane as red, and low values as blue.  Notice the uniform appearance of
   the multicolor (blue to green to red) regions, suggesting one is
   hitting instrumental noise and/or systematic effects.  Note that
   there are no regions with smaller fluctuations---no pure green
   regions.}
\label{dustTruncWholeSky}
\end{figure}

First, let us consider the effects of filtering the $l$ modes to $50 <
l < 120$.  Fig.~\ref{dustTruncBicepReg} shows the filtered Planck 353
GHz \Bmodes\ in the BICEP2 region.  Note the striations.  If we
consider the whole sky for the same map
(Fig.~\ref{dustTruncWholeSky}), we see that the pattern of striations
extends to larger scales in the low \Bmode\ regions.  The filtering
gives the map a uniformly choppy look because we are only looking at a
small range of modes.  The color scale saturates at pure red or pure
blue beyond the extremes in \Bmode\ amplitude seen in the BICEP2
region.  Fig.~\ref{dustTruncBicepReg} shows just the BICEP2 region.
It has a rainbow color scale for the amplitude of the \Bmode\ with the
red end of the spectrum positive and the blue end of the spectrum
negative.  Fig.~\ref{dustTruncWholeSky} shows this extended to the
whole sky with the map saturating at pure red or pure blue if it
exceeds the magnitude of the \Bmode\ seen anywhere in the BICEP2
region.  The modes $50 < l < 120$ are just the ones where the gravity
waves might exceed the dust in BICEP2.  At smaller scales
gravitational lensing would dominate, and at larger scales dust would
dominate because of its flatter power spectrum.  The striations appear
to cover the whole map running in different directions.  In the
galactic plane the striations have stripes parallel to the plane, but
at high latitudes they run in different directions.  Some regions look
more checkerboard like.  The regions of highest $50 < l < 120$ \Bmode\
amplitude (pure red or pure blue) follow the well-known structures in
the galaxy, where the amplitude of dust emission is greatest.  But the
regions in rainbow colors like the BICEP2 region all look surprisingly
similar, suggesting we are looking at mostly instrumental and
systematic noise.  The striations appear to be due to ringing in these
modes due to fitting the structures in high intensity regions.

The unsaturated regions (like BICEP2) look surprisingly uniform,
they all have striations that look surprisingly similar.  Once one
hits the noise level, full of ringing modes, the sky looks relatively
uniform.  There 
are no areas of lower signal (no large patches of pure green in the
map).  This suggests a large amount of noise dominating the dust
signal.  We have developed five different mapping techniques to
address these issues and produce the best ($50 < l < 120$) 353 GHz
\Bmode\ map in the BICEP2 region, with the least contamination from modes
ringing off high intensity regions elsewhere in the sky.  The BICEP2
team only had data in their survey region and would not be influenced
by modes ringing off structures elsewhere.

The first map for analysis, Map I, used the most straight-forward
means of construction of our 5 maps.  As with all maps, we took the
$a_{lm}$ modes from the {\it B} map produced by HEALPix analysis, and use
only those with $50 < l < 120$, to produce the output map.  We
further eliminated the $m = 0$ modes (in Galactic coordinates), to
dampen some of the ringing effects from the Galactic plane discussed
above.  Following the BICEP2 team's techniques, we also corrected the
map by subtracting the best-fit horizontal (in right ascension) cubic
spline, with two intervals joined in the middle of the map (which we
shall call ``desplining'').  We use this in all our maps.

For Map II, we used a similar analysis procedure.  However, we were
more aggressive in addressing the ringing in the Fourier modes from the
plane of the Galaxy.  Thus, we removed the plane by excising the
20\% of the sky centered on the Galactic Equator, and smoothly
interpolating across the region with a cosine-filter.  This technique
was quite successful in reducing the impact of the ringing, which can
be seen by the fact that the r.m.s. of the pixel values (away from
zero) is reduced from $4.97\uK$ to $3.31\uK$ in the region to be
compared with BICEP2.  The removal of the plane was sufficient
to obviate the need to remove the $m = 0$ modes.

For Map III, we reduced the ringing even further.  To do this, we
selected a mask the size of the entire BICEP2 region.  However, in
this case, we could more closely mimic the BICEP2 analysis by
desplining first, before sending the map to HEALPix.  This technique
continued to reduce the ringing noise level, leaving a standard
deviation from zero of $3.12\uK$.  Note that for these last three
maps, we used the Fortran 90 HEALPix implementation, which facilitates
use of masks.  This map is shown in Fig.~\ref{dustBLrgMask}.

For Map IV, we took our desplined map from Map III, but applied the
BICEP2 sensitivity map as the mask during the HEALPix analysis.  On
the back end we corrected for this by dividing back off by the mask to
flatten the image, just as we did for the actual BICEP2 data.  This
gives a standard deviation from zero of $2.96\uK$.  We show this map
in Fig.~\ref{dustBSensMask}.

For Map V, we did the most aggressive masking.  We took our desplined
map from Map III, and masked down to just the region of analysis
($|\mbox{RA}| \leq 30^\circ$, $-65^\circ \leq \mbox{Dec} \leq
-50^\circ$).  This ``what-you-see-is-what-you-get'' or ``wysiwyg'' map
has no influence from fitting the sky beyond the region we will actually
analyze statistically.  This gives a standard deviation from zero of
$3.03\uK$.  We regard this map to be the ``cleanest'' in the sense
that there should be minimum contamination from any other region of
the sky.

\begin{figure}
\includegraphics[width=3.25in]{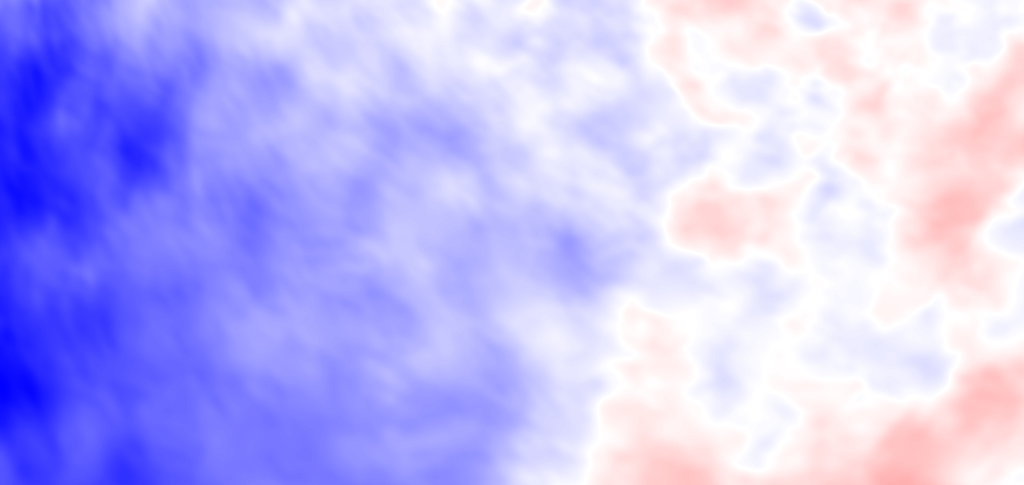}
 \caption{Planck 353 GHz polarization \Bmodes\ in Mercator projection
   in the region 
   $|\mbox{RA}| \leq 30^\circ$, $-65^\circ \leq \mbox{Dec} \leq
   -50^\circ$.  The stretch is from $-66.7\uK$ (blue) to $66.7\uK$ (red).}
\label{dustBMerc}
\end{figure}

\begin{figure}
\includegraphics[width=3.25in]{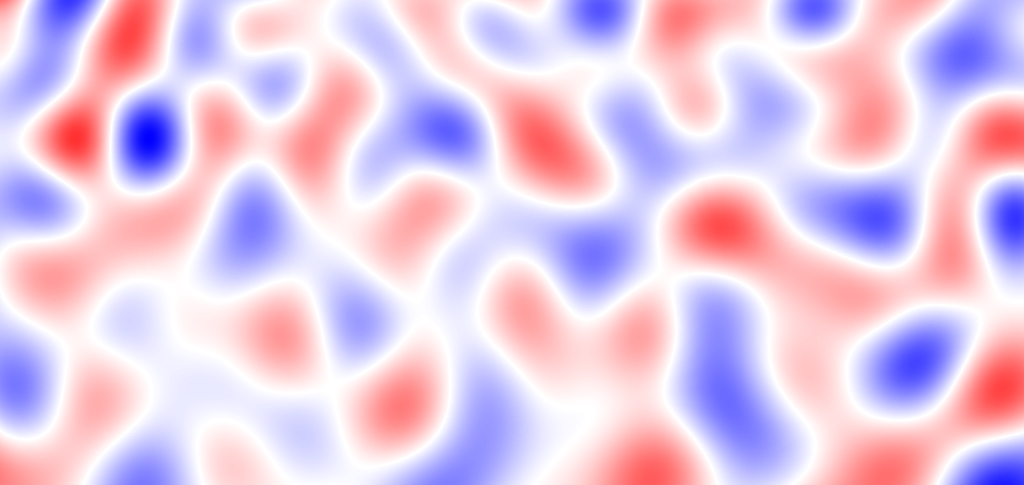}
 \caption{Map III:  The Planck 353 GHz polarization \Bmodes\ in Mercator
   projection in the region  
   $|\mbox{RA}| \leq 30^\circ$, $-65^\circ \leq \mbox{Dec} \leq
   -50^\circ$.  The map has been constructed by using only the modes
   $50 < a_{lm} < 120$ to match the BICEP2 data process.  The stretch is from
   $-11.5\uK$ (blue) to $11.5\uK$ (red).  During construction of the
   \Bmode\ map, a ``large'' mask the size of the entire BICEP2 region
   was used.}
\label{dustBLrgMask}
\end{figure}

\begin{figure}
\includegraphics[width=3.25in]{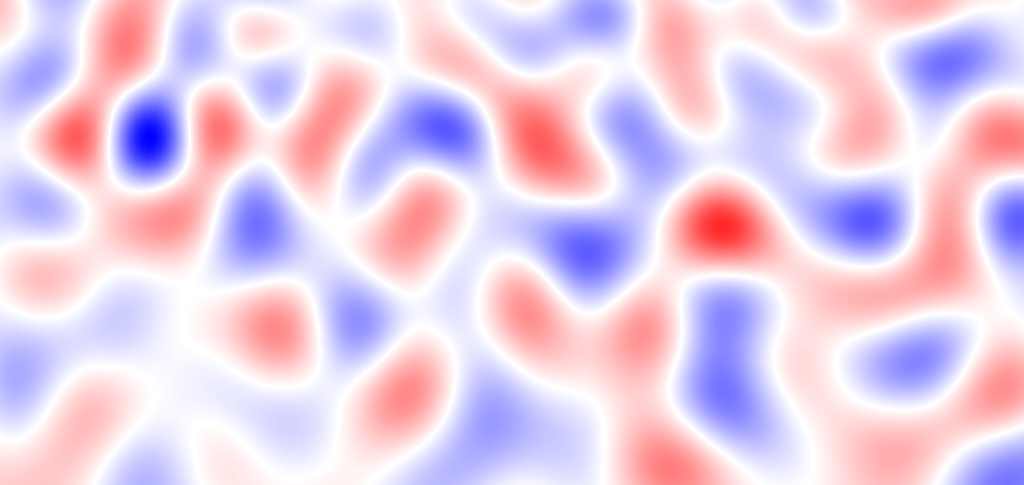}
 \caption{Map IV:  As with Map III (Fig.~\ref{dustBLrgMask}), except that
 the mask used is the BICEP2 sensitivity map itself.  This better
 approximates the apodization process BICEP2 used to reduce the
 number of ambiguous modes.  Note that there is very little difference
 between Map III and Map IV.}
\label{dustBSensMask}
\end{figure}

\section[]{Computing the Genus}

The properties of the genus are well-known in three dimensions (3D)
(Gott~\etal\ 1986, Hamilton~\etal\ 1986, Gott~\etal\ 1987), but
require some explanation in the two-dimensional (2D) case
(Melott~\etal\ 1989), particularly in the case of the sphere (Coles \&
Plionis 1991, Gott~\etal\ 1990).

We can rigorously define the 2D genus on a spherical surface (e.g.,
Colley~\etal\ 2003).  The 2D genus is defined to be equal to minus the
3D genus of solid objects formed by bestowing the regions above a
threshold with a small, but finite radial extent.  Imagine using lead
paint to paint the hot regions onto the surface of a balloon, and
after letting the paint dry, bursting the balloon to obtain solid,
curved lead shapes that would have a certain 3D genus.  Take the minus
of this number and that will be $g_{2D}$, as we will define it.

One hot spot in the north polar region would have a 2D genus of +1
(one hot spot), because the hot spot cap is one isolated region.
Suppose the hot region covered all of the sphere except for a cold
spot in the south polar region.  The genus would still be +1, because
this would look like a sugar bowl without any handles, which is also
one isolated region in 3D.  The topology in each case is identical
since one can be deformed into the other.  The genus on a plane is
determined by the local turning that a truck would do driving around the
temperature contour surface.  Circling a hot spot on a plane would
require a total turning of $2\pi$.  The Gaussian random phase formula
measures this local turning.  Circling a hot spot on the sphere
involves a total turning of $2\pi - 4\pi f$, where $f$ is the fraction
of the sphere in the hot spot (because the deficit produced by
parallel transport on the sphere is equal to the enclosed area).
Dividing by $2\pi$, we may define the effective genus:
\begin{equation}
\gtwodeff = g_{2D} - 2f,
\label{eq_g2Deff}
\end{equation}
where $f$ is the fraction of the area of the sphere in the hot spots.
For a Gaussian random phase field on the sphere
\begin{equation}
\gtwodeff \propto \nu\exp(-\nu^2/2),
\end{equation}
because the Gaussian random phase field behaves locally on the sphere
just as it does on the plane to produce this particular contribution
to the turning integral.  The Mercator projection we have chosen
preserves the turning, and thus the genus.  Therefore, in comparing
our genus curves to the random phase formula, we will use $\gtwodeff$, as
defined rigorously above (cf., Colley~\etal\ 2003).

\begin{figure}
\resizebox{\hsize}{!}{\includegraphics{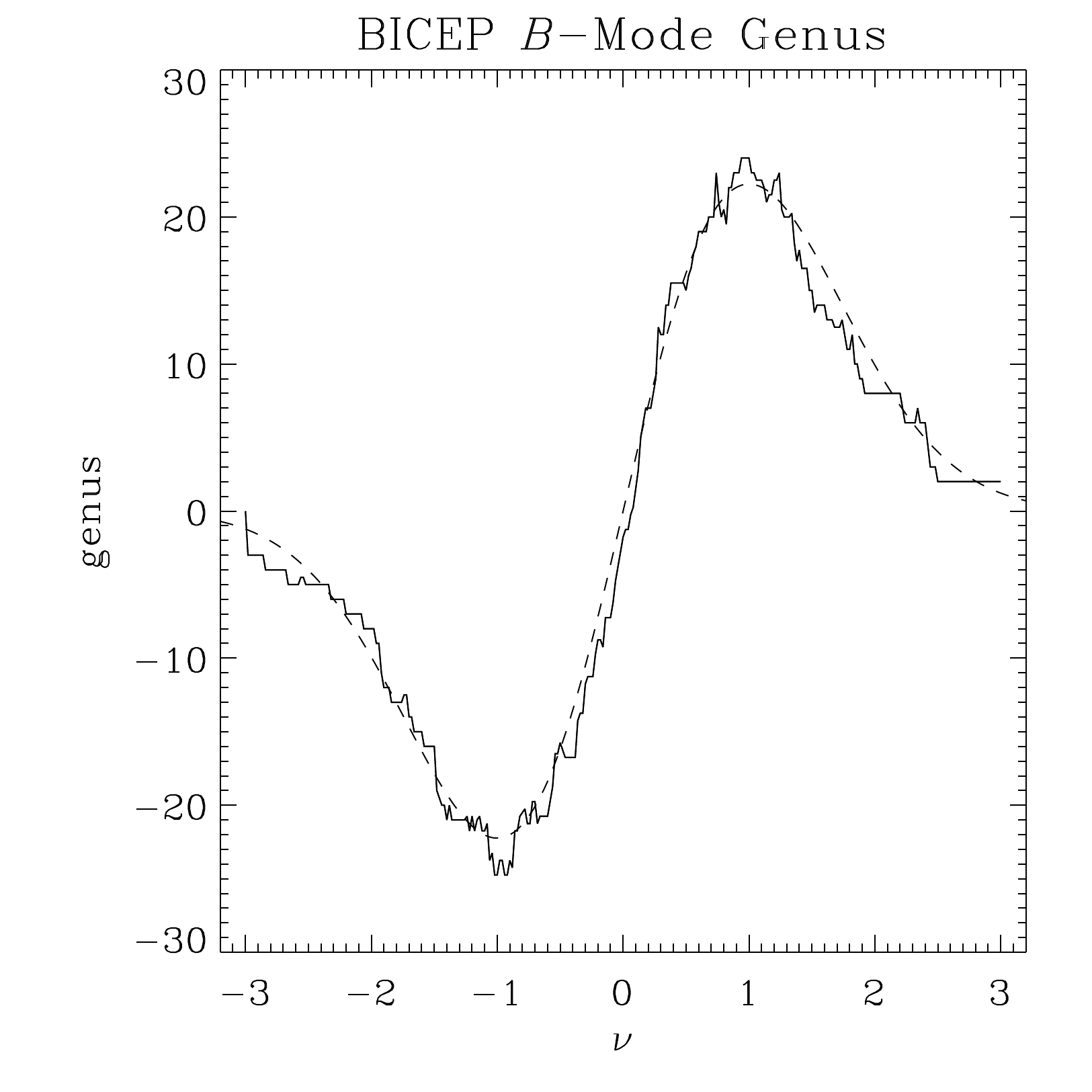}}
 \caption{2D genus for the BICEP2 data.  Overplotted is the theoretical
   Gaussian random phase genus curve, where only the amplitude has
   been fit.  The best-fit amplitude is 22.2 at $|\nu| = 1$.}
\label{bicepGenus}
\end{figure}

The BICEP2 analysis region is a fairly thin strip of the southern sky
within a fairly narrow range in declination ($-30^\circ \leq \mbox{RA}
\leq 30^\circ$, $-65^\circ \leq \mbox{Dec} \leq -50^\circ$), which
particularly lends itself to the Mercator projection (in celestial
coordinates) shown in in Fig~\ref{bicepBMerc}.  To compute the genus
we proceed with our normal two-dimensional genus calculations, using
methods very similar to those of CONTOUR2D (Melott~\etal\ 1989).
Fig.~\ref{bicepGenus} shows the 2D genus for the BICEP2 data, for a
large number of $\nu$ values.  The jaggedness of the genus trace
conveys a sense of the error in the genus computation process.  In
practice, formal error bars are difficult to estimate, because the
genus at nearby values is correlated (the same structures appear at
similar thresholds).  Gott et al. (2007) (among others) have
demonstrated elaborate techniques for estimating $\chi^2$ errors by
using a large number of simulations to create a reliable covariance
matrix that accommodates these correlations, but this is somewhat
beyond our scope here.  As a check, we did divide the region into four
quadrants so that we could estimate errorbars and found that the genus
matches the theoretical curve at ($1\sigma$) essentially across the
board (better than expected for independent variates because of the
correlations).  Overall, the fit between the BICEP2 genus and the
Gaussian random phase curve is excellent.  The BICEP2 data pass this
test.

\begin{figure}
\resizebox{\hsize}{!}{\includegraphics{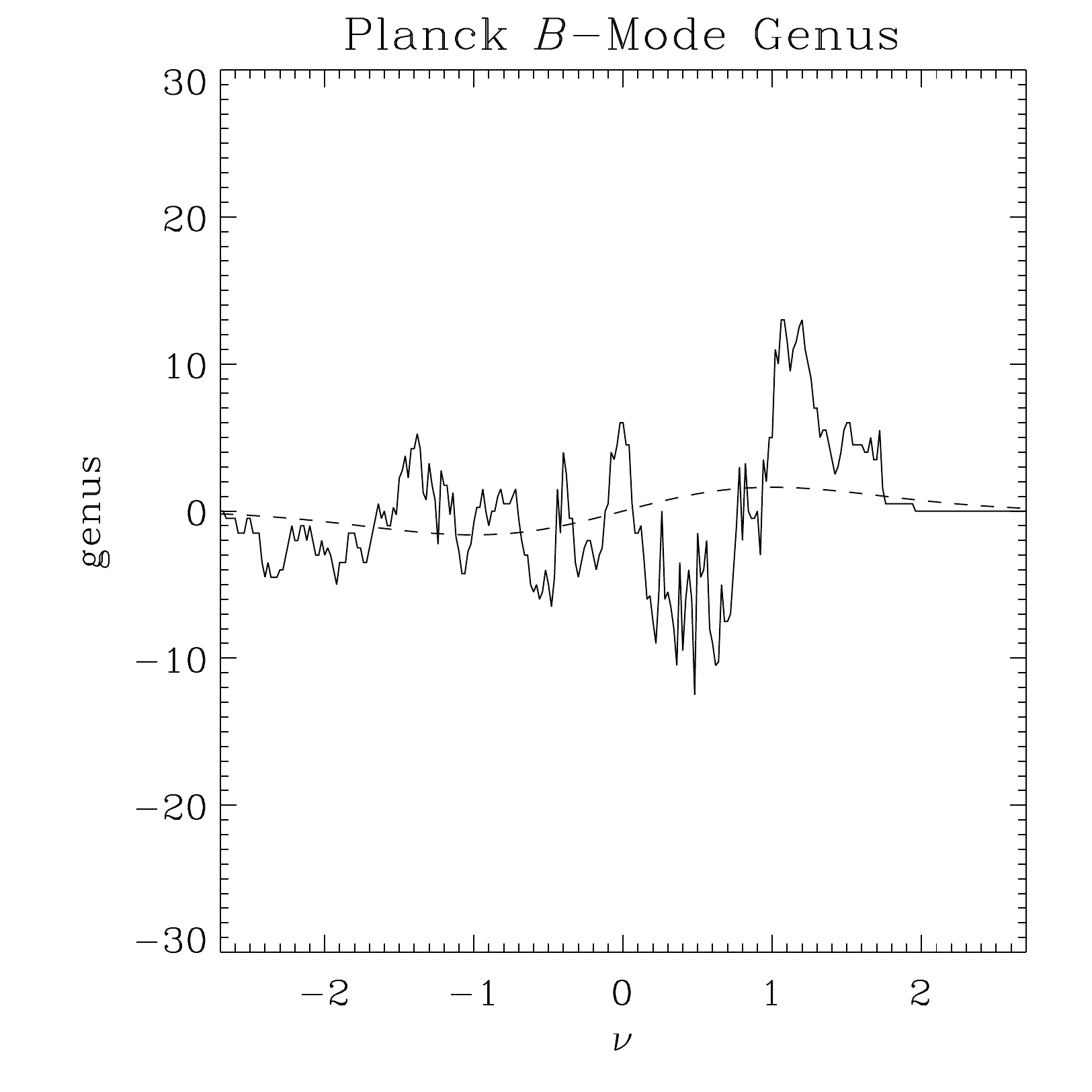}}
 \caption{2D genus for the Planck 353 GHz \Bmodes.  Overplotted is the
   theoretical 
   Gaussian random phase genus curve, where only the amplitude has
   been fit.  The best-fit amplitude is 1.62 at $|\nu| = 1$.}
\label{dustGenus}
\end{figure}

\begin{figure}
\resizebox{\hsize}{!}{\includegraphics{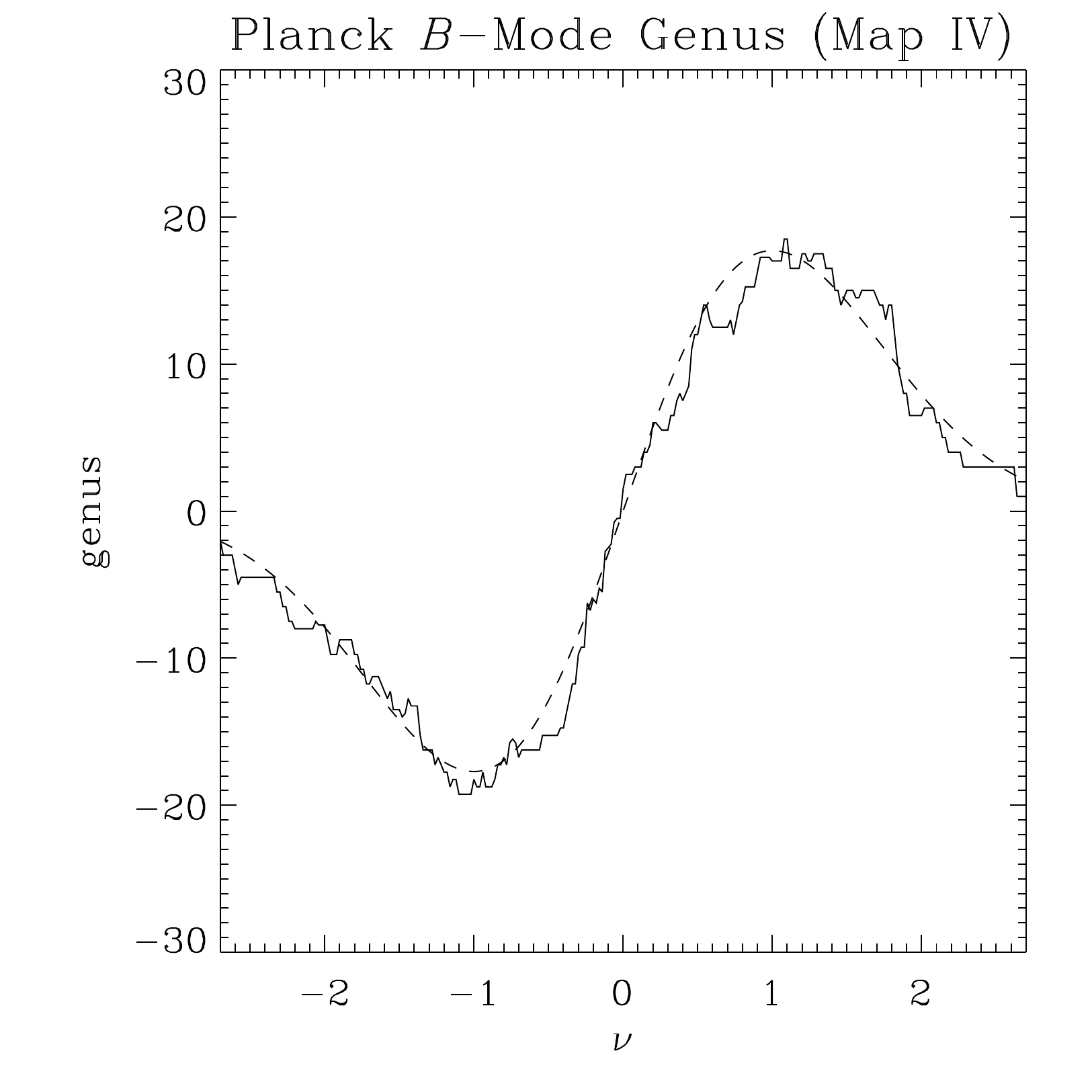}}
 \caption{2D genus for the Planck 353 GHz \Bmodes\ from our Map IV (see
   Fig.~\ref{dustBSensMask}), which has been 
   ``desplined'' in the BICEP2 region and masked to just the analysis
   region of  $|\mbox{RA}| \leq 30^\circ$, $-65^\circ \leq
 \mbox{Dec} \leq -50^\circ$, and filtered to use only the $50 < a_{lm} <
   120$ modes.  Overplotted is the theoretical Gaussian
   random phase genus curve, where only the amplitude has been fit.
   The best-fit amplitude is 17.7 at $|\nu| = 1$.}
\label{dustTruncGenus}
\end{figure}

Fig.~\ref{dustGenus} and Fig.~\ref{dustTruncGenus} provide the genus
curves for the dust maps, with all $a_{lm}'s$ and with only the $50 <
l < 120$ $a_{lm}'s$, respectively.  Not surprisingly, for the full
$a_{lm}$ dust map, which looks nothing like a Gaussian random phase
field, the genus curve looks nothing like the Gaussian random phase
theoretical curve.  However, the genus curve for the truncated
$a_{lm}$ map agrees very well with the Gaussian random phase theoretical
curve, although at lower amplitude than seen in the BICEP2 genus
curve. 

One supposes the BICEP2 team wanted to show a map that would show just
the modes where the gravity wave modes were most prominent.  The
BICEP2 team filtered out the low $l$ modes to avoid confusion with the
\Bmodes\ from gravitational lensing, and presumably filtered out the
low $l$ modes to avoid confusion with dust.  This is because the power
spectrum in the dust \Bmodes\ is very flat.  For dust $l(l +
1)C_l/2\pi \sim l^{-0.4}$.  The flat nature of the power-spectrum for
the dust is shown in Flauger, Hill \& Spergel (2014).  On the other
hand, for the \Bmodes\ expected from gravity waves over the range $10
< l < 80$ have $C_l \sim \mbox{const}$,so that $l(l + 1)C_l/2\pi \sim
l^2$.  (For $l > 100$ the gravity wave \Bmode\ spectrum begins to fall
and crosses below the gravitational lensing power at $l \approx 150$.)
The gravity wave \Bmode\ power spectrum over the range $10 < l <100$
has less power at large scales than the dust \Bmodes\ and therefore by
our genus formula a larger amplitude genus curve; the genus amplitude
is proportional to $\langle k^2\rangle$ integrated over the smoothed
power spectrum (Melott~\etal\ 1989).  The BICEP2 map has a choppier
distribution than does the Planck map.  When Flauger, Hill, and
Spergel (2014) measure the $\chi^2$ for the gravity wave fit to the
BICEP2 \Bmode\ power spectrum data ($50 < l < 175$) they obtain 1.1,
the dust fit is worse at 1.7 but not horrible.  The BICEP2 data does
follow closely the expected power spectrum for the gravity wave modes
including the fact that $l(l+1)C_l$ in the modes at $l = 50$ is lower
than at $l = 75$ as expected from gravity waves rather than higher as
would be expected for dust \Bmodes\ only.  The errors in the
individual $l$ modes are enough to make the flatter dust distribution
not look all that bad, its amplitude can be fit to give the right
level and then it is simply a bit too high at the low $l$ end and a
bit low at the high $l$ end.  This is particularly true when one is
also looking at the bump at the high $l$ end where the gravity wave
signal is flatter as well.
 
The amplitude of the genus curve can be very useful in checking the
power spectrum of the cosmological model as has been shown by
Park \& Kim (2010) and Gott et al. (2009) for 3D topology.  The fact
that Flauger, Hill, and Spergel get a better fit for the steep gravity
wave power spectrum than for the flatter dust spectrum, means that the
filtered map that BICEP2 has is in agreement with the number of
structures expected from the steep gravity wave spectrum.  This is
shown by the fact that the BICEP2 team has also included a simulation
with \Bmodes\ produced by gravitational lensing only.  The power
spectrum from gravitational lensing over the range $10 < l <100$ also
has $C_l \sim \mbox{const}$ so that $l(l + 1)C_l/2\pi \sim l^2$, it
just has an amplitude that is too low to explain the BICEP2 data as
they demonstrate.  The gravitational lensing map BICEP2 shows for
comparison has a similar number of structures, and a similar amplitude
of the genus (it just has lower contrast.)  Since the dust spectrum is
flatter (and by the way is a poorer fit to the power spectrum) it
should predict fewer structures than BICEP2 observes.  The genus curve
amplitude just provides a more dramatic illustration of this poorer
fit.

The BICEP2 {\it B} genus curve shows a maximum of 22 red spots and 22 blue
spots.  The identically filtered Planck map showed only a maximum of
18 red spots and 18 blue spots (Fig.~\ref{dustTruncGenus}).  That is
consistent with a flatter power spectrum.

\begin{figure}
\resizebox{\hsize}{!}{\includegraphics{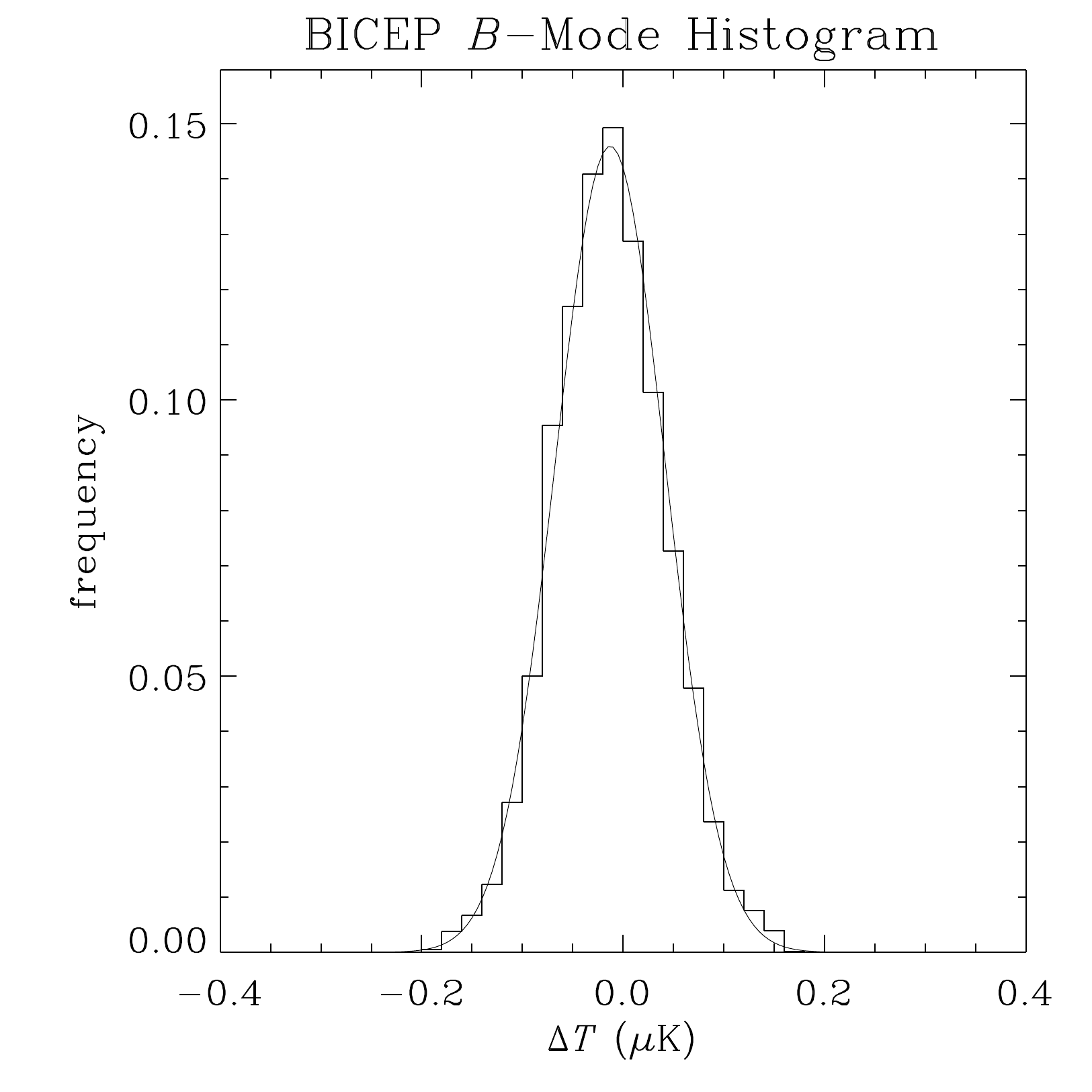}}
 \caption{Histogram from an equal-area projection of the BICEP2 \Bmode\
   data.  Overplotted is the normal curve associated with a simple
   computation of $\bar{x}$ and $s$ as estimators of the mean and
   standard deviation.}
\label{bicepHist}
\end{figure}

\begin{figure}
\resizebox{\hsize}{!}{\includegraphics{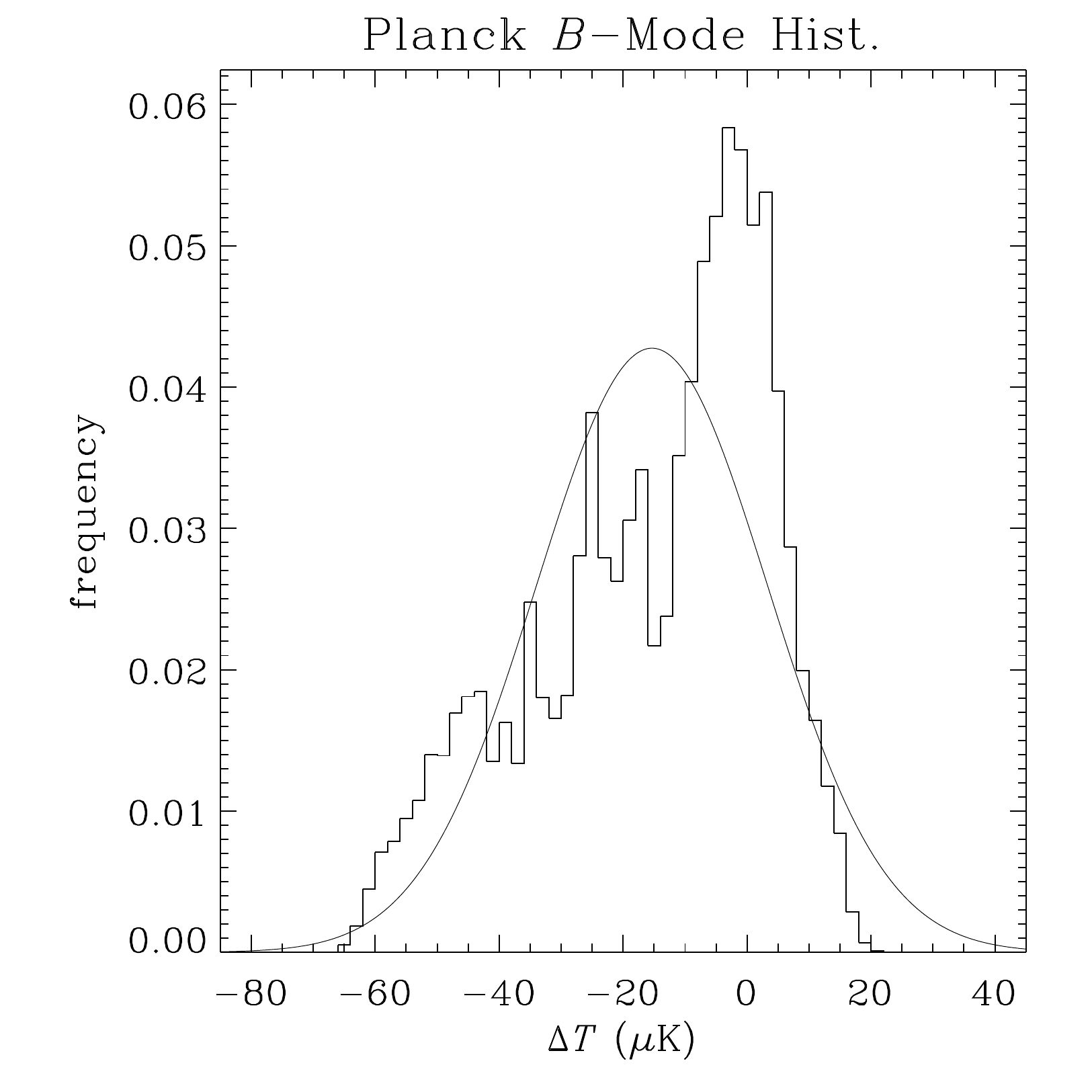}}
 \caption{Histogram from an equal-area projection of Planck 353 GHz \Bmodes,
   Overplotted is the normal curve associated with a simple 
   computation of $\bar{x}$ and $s$ as estimators of the mean and
   standard deviation.}
\label{dustHist}
\end{figure}

\begin{figure}
\resizebox{\hsize}{!}{\includegraphics{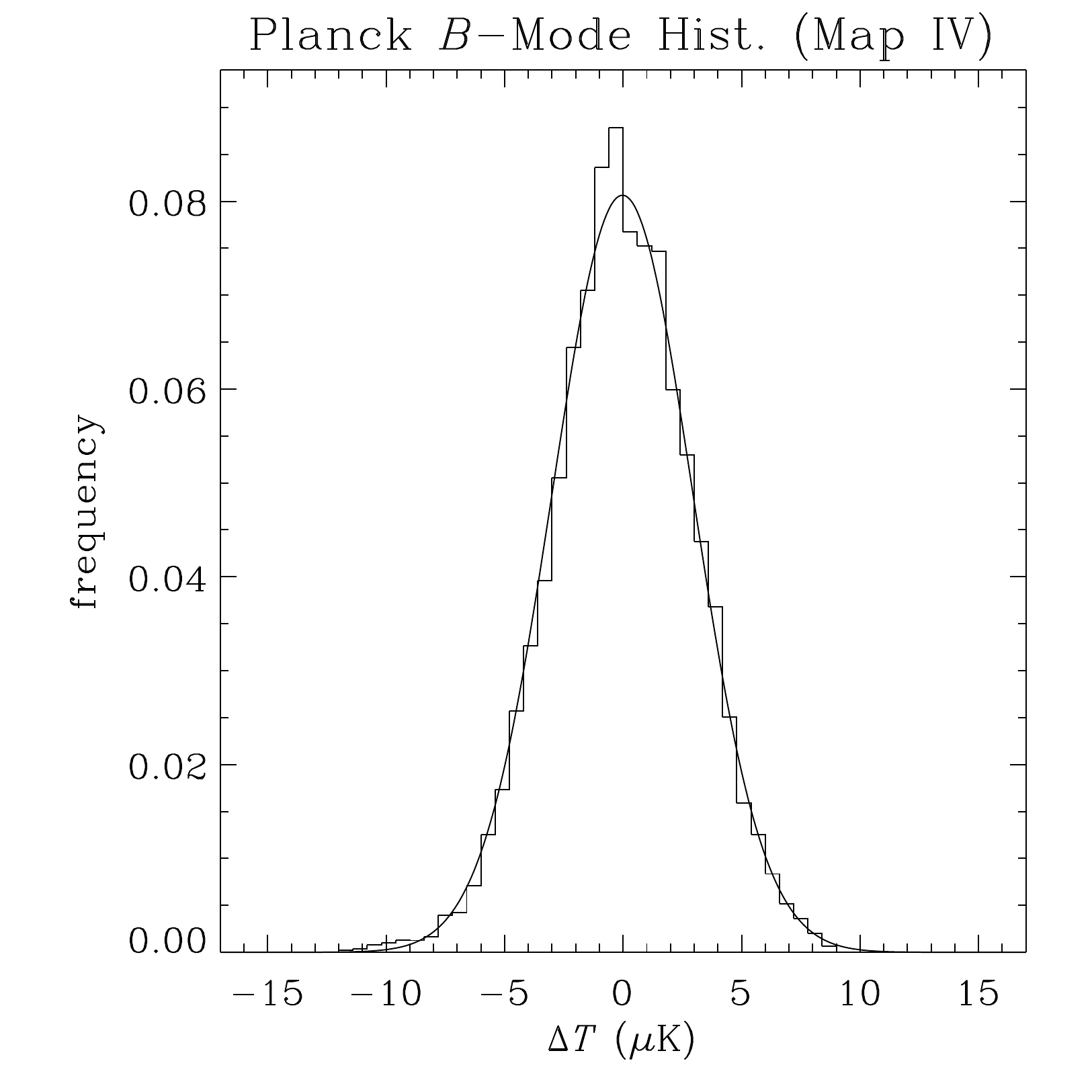}}
 \caption{Histogram from an equal-area projection of Planck 353 GHz \Bmodes,
   computed after ``desplining'' in right ascension;  only $50 <
   a_{lm} < 120$ modes are included (Map IV). 
   Overplotted is the normal curve associated with a simple 
   computation of $\bar{x}$ and $s$ as estimators of the mean and
   standard deviation.}
\label{dustTruncHist}
\end{figure}

For a Gaussian random phase field, one also expects a Gaussian
histogram.  To calculate this we constructed maps using the Lambert
equal area cylindrical projection (not shown), rather than the
Mercator projection; this histogram for the BICEP2 \Bmodes\ is shown
in Fig.~\ref{bicepHist}.  The Gaussian with $\mu \approx \bar{x}$ and
$\sigma \approx s$ is shown on top of the histogram.  As with the
genus, the fit is excellent.  This distribution is consistent with the
Gaussian Random phase distribution expected from a cosmological origin
due to gravity waves from inflation.  We also show the histograms for
the Planck maps, and as one might expect from Figs.~\ref{dustBMerc},
\ref{dustBLrgMask} and \ref{dustBSensMask}, the full $a_{lm}$ map shows
a highly non-Gaussian distribution (Fig.~\ref{dustHist}), while the
truncated $a_{lm}$ map shows a distribution very much consistent with
a Gaussian (Fig.~\ref{dustTruncHist}).  As such, the BICEP2-like
processing of the data appears to mask many non-Gaussianities in the
dust-dominated Planck map.  Kamionkowski \& Kovetz (2014) have
suggested using the hexadecapolar departure from isotropy to
reveal the level of non-Gaussianity in the \Bmode\ data introduced by
foregrounds, but conclude the current signal-to-noise is insufficient
for conclusive results.

\section{Discussion}

If $r = 0$ and all the \Bmodes\ seen in the BICEP2 map are produced by
dust, and Planck is detecting dust modes with ($50 < l < 120$) at high
signal-to-noise then we would expect the two maps
(Fig.~\ref{bicepBMerc} and Fig.~\ref{dustBSensMask}) to have the same
number of structures in the same locations.  Visual inspection
confirms the genus result that there are indeed fewer structures in
the Planck map.  If Planck is detecting the dust modes this suggests
that a number structures in the BICEP2 map must be due to gravity
waves.  The 353 GHz map itself, of course, contains more information
than contained in the power spectrum or the histogram or genus curve.
We can ask for the correlation coefficient between the two maps.  It
is 15.2\% (average for Maps I--V) showing that the location of the
structures are mostly uncorrelated---as is apparent from a visual
inspection.  The \Bmodes\ that BICEP2 is seeing are not the ones we are
seeing in the Planck 353 GHz map.

The \Bmodes\ in BICEP2 are of too large an amplitude to be produced by
gravitational lensing, as shown by a simulation which includes a
standard flat lambda model and gravitational lensing only (BICEP2
Collaboration 2014a).  The \Bmodes\ due to gravity waves also have $l(l
+ 1)C_l/4\pi \sim l^2$ at low $l$ but are of lower amplitude.  This
simulated map has a similar number of structures (consistent with what
we would expect from the genus statistic), but its amplitude is too
small to fit the BICEP2 data.  This aspect of their study has not been
questioned.  In addition BICEP2 at smaller scales sees the
gravitational lensing \Bmodes.  The gravity wave modes kick in for $l
< 100$ with higher amplitude in the \Bmode\ power spectra, over and
above the gravitational lensing \Bmodes.  So it seems clear that
BICEP2 is not seeing just gravitational lensing.

At this point we should mention another way to produce
\Bmodes\---Faraday Rotation---in order to immediately rule it out.
Imagine that one has a pure \Emode\ with polarization directions that
radially point away from a center.  Faraday rotate all these
polarization directions by $45^\circ$ and you have produced a pure
\Bmode\ with a pinwheel pattern.  The maximum amplitude of the \Emode\
detected by BICEP2 at $50 < l < 120$ is $\sim 1.7\uK$ while the
maximum amplitude of the \Bmode\ is $\sim 0.3 \uK$ To produce this much
\Bmode\ from an \Emode\ would require a Faraday rotation of $\theta \sim
\tan^{-1}[(0.3/\sqrt{2})/(1.7 + 0.3/\sqrt{2})] \sim 6.3\deg$.  Now the
galactic foreground Faraday rotation in the BICEP2 region is fairly
uniform at about $100 \mbox{rad}\cdot m^{-2}\lambda^2$
(Opperman~\etal\ 2011).  At 150 GHz, $\lambda = 0.002\mbox{m}$, so the
Faraday rotation is 0.0004 rad, far less than the required 0.1 radian.
The additional Faraday rotation seen in quasars amounts to only an
additional $\pm50\mbox{rad}\cdot m^{-2}\lambda^2$ due to intervening
clouds.  Faraday rotation near recombination also is implausible since
the wavelengths are shorter by a factor of 1000 then which lowers the
Faraday rotation by a factor of a million.  In any case, if Faraday
rotation were responsible for the \Bmodes\ the \Bmode\ and the \Emodes\
in the BICEP2 map would be at identical locations---this is not the
case. The {\it E-} and \Bmodes\ are at different locations in the BICEP2 maps.
Thus, Faraday rotation can be ruled out as the source of the BICEP2
\Bmodes.

Flauger, Hill and Spergel (2014) have stated that the polarized \Bmode\
amplitude is sufficiently uncertain that the results are either
consistent with a \Bmode\ signal from BICEP2 due to $r = 0.2$ and
gravity waves or $r = 0$ and due to dust.  We have not depended on the
amplitude at all but have only gone on the structure of the 353 GHz
modes, which do not match the visual appearance nor the amplitude of
the genus curve in the BICEP2 data.  If the structures seen in the
BICEP2 map could readily identified with structures in the 353 GHz
Planck map this would be a smoking gun implicating the dust.  This is
definitely not seen.

\subsection{Amplitude Considerations}

Now let us look at the amplitude for the first time.  Our histogram
shows the amplitude of the \Bmodes\ amplitude in the (spline
subtracted) filtered $50 < l < 120$ BICEP2 map to be $0.0838\uK$
($1\sigma$) at 150 GHz.  The histogram of the \Bmode\ amplitudes in
the 353 GHz Map IV from Planck which we have constructed from their
{\it Q} and {\it U} maps show the (spline subtracted) filtered $50 < l
< 120$ \Bmode\ amplitude to be $2.96\uK$, again in brightness
temperature.

We have measured the power spectrum of our whole sky \Bmode\ dust map
at 353 GHz; it follows the simple power law form found by Planck and
is at a higher amplitude than their 80\% of the sky map which avoids
the galactic plane.  To compare with that 80\% map, we have cleaned
the Planck map of the 20\% of the sky that includes the plane and we
get the same amplitude power spectrum that the Planck team reports
(Boulanger~\etal\ 2014).  This makes it clear that the scale in the
Planck {\it Q} and {\it U} maps is in units of K in brightness
temperature, which is what we have adopted.

The Planck collaboration estimates that the dust \Bmode\ power spectrum should
be lowered by a factor of $25.8^2$ at 143 GHz relative to that at 353
GHz.  We have taken this value from a power spectrum estimate for 143
GHz they have made publicly available.  That implies they expect the
amplitude of the dust signal at 143 GHz relative to 353 GHz to be
lower by a factor of 25.8.  If dust emission goes as
$I_\nu \sim \nu^\gamma$ over that frequency range and the polarization fraction
stays constant between 143 GHz and 353 GHz, then the amplitude of the
dust map at 143 GHz should be lower (in brightness temperature) than
that in the 353 GHz map by a factor of
\begin{equation}
{{e^{143/56.8}}\over{e^{353/56.8}}}
\left[{{e^{353/56.8}-1}\over{e^{143/56.8}-1}}\right]^2
\left[{{353}\over{143}}\right]^{\gamma-4} = 25.8,
\end{equation}
where for the microwave background, $h\nu = kT$ at 56.8 GHz.  This implies
a value of $\gamma = 3.324$ over the frequency range from 143 GHz to 353 GHz
which is plausible.  If we use that value we can calculate using the
same formula the lowering we expect from 353 GHz to 150 GHz.  We find
a factor of 23.1, (quite reasonable since we are moving over a
slightly smaller frequency range).

Probably the best estimate of this factor is from the recent Planck
data (Planck Collaboration 2014a) on polarization as a
function of frequency.  They estimate that the polarized dust emission
$I_\nu \sim \nu^{3 + \beta}/[e^{h\nu/kT_D} -1]$, where $\beta = 1.65$,
and $T_D = 19.8\mbox{K}$ for high latitude dust.  (Thus $h\nu/kT_D =
\nu/413.5\mbox{GHz}$.)  This includes the fact that the polarized
fraction is slightly lower at 150 GHz than at 353 GHz.  This is
consistent with Draine and Hensley's (2013) treatment of ferromagnetic
nanoparticles in interstellar dust.  For inclusions, magnetic dipole
emission is expected to be polarized orthogonally relative to the
normal electric dipole radiation.  This can explain a somewhat smaller
polarization fraction at 150 GHz than at 353 GHz.  The factor $\beta$
is derived empirically by the Planck team.  We find that for polarized
emission
\begin{equation}
I_{353}/I_{150} = \left({{353}\over{150}}\right)^{4.65} {{e^{150/413.5} -
    1}\over{e^{353/413.5} - 1}} = 17.35.
\end{equation}
Converting fluctuations in intensity to fluctuations in brightness
temperature in the CMB using 
\begin{equation}
dI = d\left[{{2\pi h\nu^3c^{-2}}\over{e^{h\nu/kT} - 1}}\right] =
{{2\pi h\nu^3c^{-2}e^{h\nu/kT}}\over{(e^{h\nu/kT} - 1)^2}}(h\nu/kT^2)dT,
\end{equation}
where $h\nu/kT = h\nu/(k\cdot 2.72\mbox{K}) = \nu/56.8\mbox{GHz}$, we find
for polarized dust emission:
\begin{eqnarray*}
{{\Delta T_{353}}\over{\Delta T_{150}}} = 
17.35\cdot
\left[{{353}\over{150}}\right]^{-4}\left[{{e^{150/56.8}}\over{e^{353/56.8}}}\right]
\left[{{e^{353/56.8} - 1}\over{e^{150/56.8} - 1}}\right]^2\\
 = 21.3,
\end{eqnarray*}
which we will adopt.

This is quite similar to the original Planck derived estimate of 23.1
mentioned above, but this is more accurate, based on later and more
complete data.

As we shall see later, the BICEP2 map and the Planck 353
GHZ map have a correlation of only 15.2\%.  If we attributed that to
supposing that the dust signal at 150 GHz was 15.2\% of the total
BICEP2 signal which we would believe was primarily due to gravity
waves, we would be left with the untenable conclusion that the
polarized dust emission in this particular region must fall off as we
go from 353 GHz to 150 GHz by a factor of 11.1 {\it more} than we
expect (i.e., a factor of [$0.142\uK/0.0838\uK]/0.152$).  Dust can
have different polarization fractions at different frequencies due to
grain properties, but this much lowering seems implausible.  In
addition the Planck data shows evidence for a small change in
polarization fraction from 150 GHz to 353 GHz and this has already
been included.  So if the Planck 353 GHz map were a pure dust signal
with little noise contamination, the scenario that the dust signal at
150 GHz was sub-dominant would not work.

Can the fallen-off dust signal be equal to the BICEP2 signal?  Against
that simple interpretation (that the BICEP2 signal is a pure high
signal-to-noise dust signal which can be seen in the 353 GHz map) is
the fact that the observed BICEP2 pattern does not match the Planck
353 GHz pattern.

By the way, we expect the BICEP2 $50 < l < 120 B$ mode signal if it is
cosmological and due to gravity waves, to have the same amplitude in
brightness temperature at 353 GHz as it does at 150 GHz: $0.0838\uK$
($1\sigma$).  Thus, under no circumstance do we expect the gravity
wave signal to significantly pollute the Planck signal of $2.96\uK$
($1\sigma$) at 353 GHz.

By contrast we may expect the dust emission at 353 GHz to pollute the
BICEP2 signal at 150 GHz to some extent.  Indeed this seems to be the
case.  We do find a correlation coefficient of 0.181 between the
filtered BICEP2 map and the similarly filtered 353 GHz map from Planck
(Map IV, shown in Fig.~\ref{dustBSensMask}).  This suggests that the
dust is peeking through in both maps (the correlation is positive).

We now consider the uncertainties associated with the correlation
coefficients $C_{PB}$.  To this end, for each map, we computed the
correlations between not only the dust map and the BICEP2 map, but
also between the vertically and/or horizontally flipped dust map and
the BICEP2 map.  This gave us 3 fairly independent measures for the
level of correlation one might expect if there were no dust
contamination.  We went one step further, which was to conduct exactly
the same experiment using maps constructed from the sky at the
opposite galactic longitude, where there appears to be very similar
structure in Fig.~\ref{dustTruncWholeSky} to that in the BICEP2
region.  Each of the four available flips should be uncorrelated
with the BICEP2 map.  This gives us 7 presumably uncorrelated maps to
evaluate for each of our map methods.  With our 5 methods, we now have
35 values from which to form an estimate of the typical variation in
the correlation.  The standard deviation for all 35 is
$0.039$.  This is consistent with the variation seen in the last 4
maps (Maps II -- V, where the impact of the plane has been reduced one
way or the other).  We therefore regard this to be a reasonable
errorbar on each of our correlation results.  If we simply take the
mean of the 5 correlation values, we have $0.152$, which is more than
three standard deviations away from zero.  We therefore regard the
correlation to be significantly detected.

Let us consider the possibility that the Planck signal in the ($50 < l
< 120$) modes is significantly polluted by noise and/or systematic 
effects.  If the Planck map were all noise, how would one explain the
correlation with BICEP2?  But some noise contamination of the Planck
$50 < l < 120$ map could help explain why its genus curve and
histogram approximate the Gaussian random phase results as well as
they do.

\subsection{Calculations of the contributions of various components}

\begin{figure}
\includegraphics[width=3.25in]{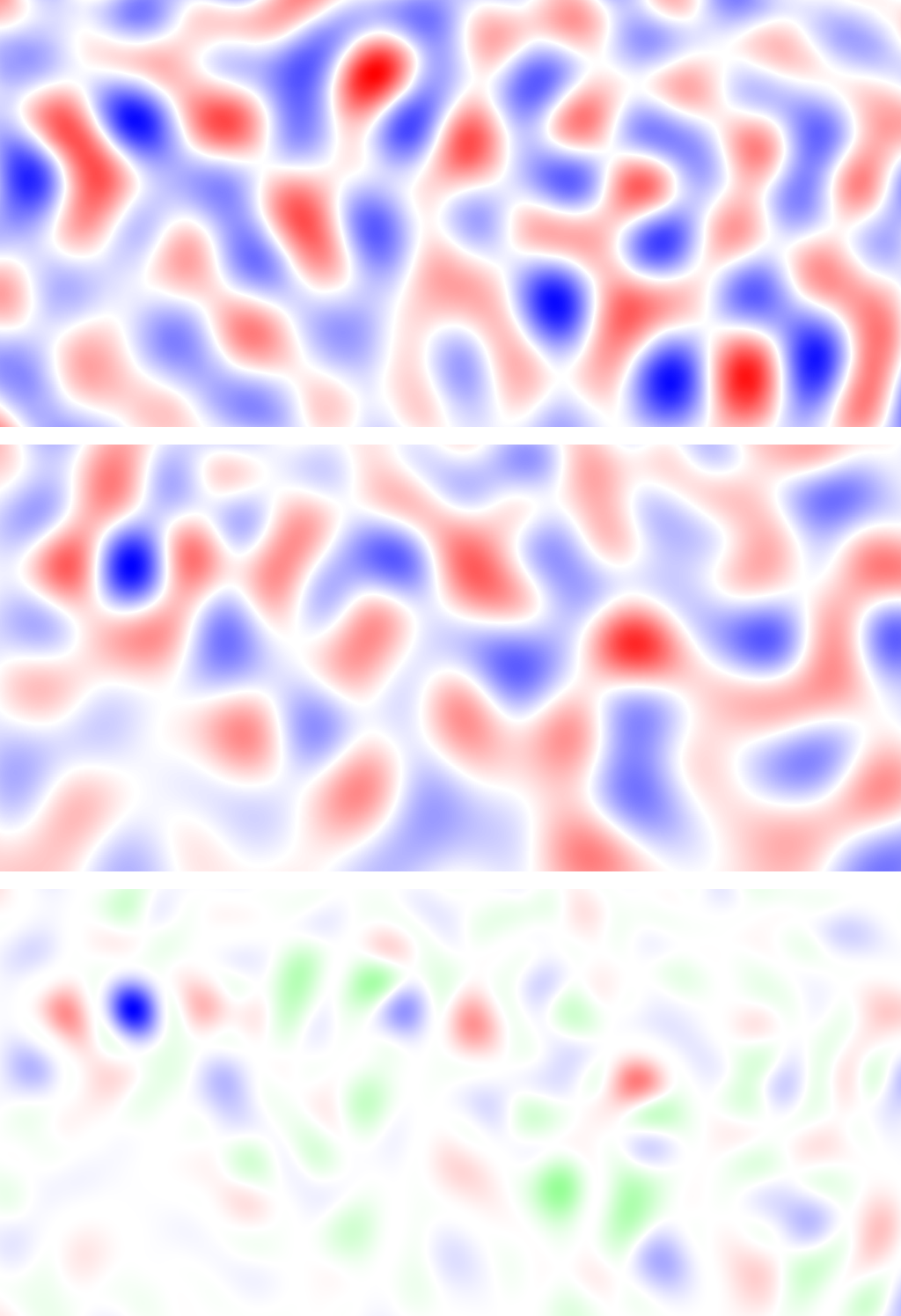}
 \caption{At top, the BICEP2 map (as in Fig.~\ref{bicepBMerc}); in the middle,
   our Planck 353 GHz Map IV (as in Fig.~\ref{dustBSensMask}).  At bottom is the
   correlation of these two maps.  All maps are in Mercator
   projection in the region 
   $|\mbox{RA}| \leq 30^\circ$, $-65^\circ \leq \mbox{Dec} \leq
   -50^\circ$.  Red shows positive-positive correlations; blue shows
   negative-negative correlations; green shows anti-correlations
   (negative-positive or positive-negative).}
\label{corrRGB}
\end{figure}

Now we will analyze the situation in detail.  We have
\begin{equation}
\sigma_B^2 = \sigma_{BGW}^2 + \sigma_{BN}^2 + \sigma_{BGL}^2 +
\sigma_{BD}^2,
\end{equation}
where $\sigma_B$ is the standard deviation in the BICEP2 map,
$\sigma_{BGW}$ is the standard deviation of the BICEP2 gravity wave
signal, $\sigma_{BN}$ is the standard deviation of the BICEP2 noise,
$\sigma_{BGL}$ is the standard deviation of BICEP2 gravitational
lensing signal, and $\sigma_{BD}$ is the standard deviation of the
BICEP2 dust signal (since all these are uncorrelated with each other,
they add in quadrature to give $\sigma_B^2$ for the whole map).
The BICEP2 team has produced a simulated map showing only the expected
gravitational lensing and noise.  From our digitization of this map,
which includes only gravitational lensing and noise, we find its
standard deviation to be $\sigma_{sim} = 0.0561\uK = 0.670\sigma_B$
(somewhat higher than the approximate value of 0.5 that the
BICEP2 team implied in their figure.  To be precise, they said that
their map amplitude was more than twice the simulation amplitude at
the $l = 70$ mode [we verified this].  That is in the middle of the
range where the 
gravity waves are most prominent, so the overall ratio should be
expected to be somewhat less, which is what our digitization shows.).
The simulation has an amplitude
\begin{equation}
\sigma_{sim}^2 = \sigma_{BN}^2 + \sigma_{BGL}^2 = [0.0561\uK]^2 =
0.448 \sigma_B^2.
\end{equation}
We now consider the correlation between the dust map and the BICEP2
map.  In Fig.~\ref{corrRGB} we show the correlation by shading in red
positive-positive correlations, in blue negative-negative correlations
and in green negative-positive or positive-negative anti-correlations.
One can see that the most significant features in the dust map (one
major red spot and one major blue spot) are, in fact, correlated with
the BICEP2 map.  To calculate the correlation coefficient $C_{PB}$
between the Planck and BICEP2 maps, we divide each map by its standard
deviation, and then multiply the two maps and average over the pixels.
The noise in the BICEP2 and the noise in the Planck are uncorrelated,
and so their product averages to zero.  The signal terms appear as
products (e.g., $\sigma_{BGW}\sigma_{PGW} = \sigma_{BGW}^2$).  The
gravity wave and gravitational lensing signals should be equal in both
maps.  We will now write the formula for the Planck 353-BICEP2
correlation coefficient, $C_{PB}$:
\begin{equation}
C_{PB} = {{\sigma_{BGW}^2 + \sigma_{BGL}^2 +
    \sigma_{BD}\sigma_{353D}}\over
{\sigma_B\sigma_{353}}}
\end{equation}

Since the noise in BICEP2 and noise in Planck are uncorrelated, and
since $\sigma_{BGW} = \sigma_{PGW}$ and $\sigma_{BGL} = \sigma_{PGL}$, we know
$\sigma_{BN}^2 + \sigma_{BGL}^2 = 0.448 \sigma_B^2$, but not the
specific value of $\sigma_{BGL}^2$. So let's set $z =
\sigma_{BGL}^2/\sigma_B^2$.  Then $0 < z < 0.448$.  Set $w =
\sigma_N^2/\sigma_B^2$.  Then $w + z = 0.448 = z_{max}$.  From our
calculation of the frequency effects on the dust signal, we have
$\sigma_{353D} = 21.3\sigma_{BD}$.  Substituting, we get
\begin{equation}
C_{PB}\sigma_{353} = {{\sigma_{BGW}^2 + \sigma_{BGL}^2 +
    21.3\sigma_{BD}^2}\over{\sigma_B}}
\end{equation}
With $x = \sigma_{BGW}^2/\sigma_B^2$
and $y = \sigma_{BD}^2/ \sigma_B^2$ we have
\begin{equation}
C_{PB}\sigma_{353}/\sigma_B = x + z + 21.3y
\end{equation}
Referring to equation (6) and (7), we find 
\begin{equation}
x + y = 1 - \left[(\sigma_{BN}^2 + \sigma_{BGL}^2)/\sigma_B^2\right] = 1 - 0.448 = 0.552
\end{equation}
We can estimate the value of $z$ from Fig.~2 of the BICEP2 paper
(BICEP2 2014a), which shows their measured power spectra, and
theoretical contribution due to gravitational lensing.  We know the
noise power spectrum $C_l \sim \mbox{const}$, just as with the gravitational lensing power spectrum, and we can determine its
amplitude from the stated value of $87\mbox{nK}$ in a square degree patch.
Using HEALPix, we construct a $C_l \sim \mbox{const}$ map, and measure
the rms amplitude in 1 square degree patches.  We can measure the
theoretical $C_l$ for gravitational lensing at $l = 119$ from Fig. 2
in the BICEP2 paper (BICEP2 2014a).  The theoretical $C_l$ amplitude
is of course based on lensing data from Planck and elsewhere.  Taking
the ratio of the $C_l$'s from lensing and noise at $l = 119$ allows
us to calculate that the ratio $z/w = 0.774$.  The noise and
gravitational lensing power spectra are proportional to each other
(both have $C_l \sim \mbox{const}$ over the range $50 < C_l < 120$), so
we can get the ratio $z/w$ from the ratios of the $C_l$'s at $l =
119$.  We will therefore adopt $z/w = 0.774$.  Then, using $w + z =
0.448$, we find $z = 0.1955$.  We can now solve the two equations above for $x$
and $y$.  Substituting we find
\begin{equation}
C_{PB}(\sigma_{353}/\sigma_B) = 0.552 + 0.1955 + 20.3y
\end{equation}
We can then solve for $y$ and find $x = 0.552 - y$.  The equation can
also be rewritten as
\begin{equation}
C_{PB}(\sigma_{353}/\sigma_B) = 1 - w + 20.3y,
\end{equation}
which we will find useful later.  To estimate the tensor-to-scalar
mode ratio $r$, we simply utilize the BICEP2 team's power spectrum
calibration.  In our notation, that is simply $r = (x/0.552) \cdot
0.2$.  In other words, if there were no dust ($y = 0$), $x = 0.552$
and we would find $r = 0.2$, as calibrated by BICEP2's power spectrum
analysis.  We present in Table~\ref{xyzrTable} those results for Planck
Maps I -- V.
\begin{table}
\caption{Relative contributions to BICEP map, for different analyses
of the dust.  The error on each measured correlation, $C_{PB}$, is
estimated to be 0.039.  $C^*$ refers to the correlation necessary to
imply a $r$ of 0.  In all cases, the correlation measured is at
least $2.5\sigma$ below this level (for Maps I--V, these levels are
$2.6\sigma$, $4.9\sigma$, $3.1\sigma$, $4.1\sigma$ and $3.1\sigma$).} 
\label{amplitudes}
\begin{tabular}{lcccccc}
\hline
Map & $\sigma_{353}(\uK)$ & $C_{PB}$ & $C^*$ & $x$ & $y$ & $r$\\
\hline
Map I & 4.97 & 0.101 & 0.202 & 0.292 & 0.259 & 0.106 \\
Map II & 3.31 & 0.112 & 0.302 & 0.370 & 0.181 & 0.134 \\
Map III & 3.12 & 0.202 & 0.321 & 0.219 & 0.333 & 0.079 \\
Map IV & 2.96 & 0.181 & 0.337 & 0.274 & 0.278 & 0.099 \\
Map V & 3.03 & 0.161 & 0.331 & 0.301 & 0.250 & 0.109 \\
\hline
\end{tabular}
\label{xyzrTable}
\end{table}

The mean value of $r$ from the 5 methods is $r = 0.106$.  A simple
estimate of the uncertainty associated with our values of $r$ is the
direct standard deviation of the above values from the different
methods; that computes to $\pm 0.020$.  This is a reasonable estimate
of the error associated with our varied mapping processes.  (For
comparison, one could apply median statistics [c.f. Gott~\etal\ 2001]
to our 5 $r$ values; the median value is $r = 0.106$, while the chance
the true value is between 0.99 and 0.109 is 62.5\%, roughly
$1\sigma$.)  If we used the independent $\sigma_{353D} =
23.1\sigma_{BD}$ estimate from a simple power-law interpolation
between Planck at 353 GHz and Planck at 143 GHz to estimate the dust
amplitude at 150 GHz, we would have gotten a mean value of $r = 11.4$.
Thus the uncertainty in $r$ to due the uncertainty in this ratio is
$\pm 0.008$.  However, there is still some additional error in the
estimate of $z$.  The BICEP2 team reports that the gravitational
lensing power can vary by about 45.5\%.  As such, we recomputed our
$x$, $y$ and $r$ values with $z$ increased and decreased by 45.5\%
($z_{max}$ was adjusted by the same resulting addends on $z$); this
introduces an additional error of $0.016$.  Note that Equation (12)
shows that adding or subtracting ($0.455 \times 0.1955$) = 0.089 from
$z$ changes $y$ not at all, but adds or subtracts $0.089$ from $x$
with consequent changes of $r$ of $\pm 0.016$.  We also have the error
in $r$ introduced by the $0.039$ error in the correlation
measurements; this translates to an error of $0.029$ in $r$.  Each
individual map has an uncertainty in its correlation coefficient
$C_{PB}$ of $\pm 0.039$ determined as we have described, by
cross-correlating BICEP2 with random Planck 353 fields.  We raise and
lower $C_{PB}$ by this amount to compute the error bias on $r$ in each
of the 5 maps.  The rms value of this $1\sigma$ error in $r$ is 0.029.
So, we take as our best value $r = 0.106$ (this is both the mean and
the median of the values from our 5 maps).  As our very conservative
estimate of the error in $r$, we will add in quadrature the standard
deviations of the $r$ values from the 5 different maps, errors in the
factor 21.3, the errors due to the expected errors in the correlation
coefficients, and the errors due to the uncertainty in gravitational
lensing: $r = 0.106 \pm 0.039$.  Rounding and keeping significant
digits, $r = 0.11 \pm 0.04$.

It is important to note that these varied methods give consistent
results.  Map I, for example, includes ringing in the $50 < l < 120$
modes from the Galactic plane.  This ringing just adds noise, which
boosts the value of $\sigma_{353}$ and lowers the correlation by a
factor of 1.7 relative to the lower noise Map IV, but both give similar
values of $r$.

Furthermore, we can explore what value of the correlation would be
necessary to drive the gravitational wave component, $x$ to zero in
our maps (see Table~\ref{xyzrTable}).  For Map IV, we find that a
correlation of 33.7\% would be necessary.  This value is more than
$3\sigma$ outside our observed correlation of $0.181 \pm 0.039$.

Of course, the $\sigma$ value from the BICEP2 map is somewhat lower
than one might na\"ively expect from the power spectra given by the
BICEP2 team, due to desplining and limiting the spherical harmonic
modes to $50 < l < 120$.  We find for a pure noise map, for the Planck
map, and for the BICEP2 map, the power spectra were all suppressed
by an equivalent factor by all of this processing, which leaves the
analysis of the ratios and correlations intact.

As an example, suppose the BICEP2 analysis drops ambiguous modes (to
avoid leakage of \Emodes\ into \Bmodes) and drops other modes for
experimental reasons.  Their calculation of the power spectrum will
take these drop-outs into consideration, but the map might be missing
these modes.  Thus $\sigma_{B}^2 = \sigma_1^2 + \sigma_2^2$, where the
map includes modes labeled 1 and excludes the modes labeled 2.  The
observed BICEP2 map will have an amplitude $\sigma_1$, slightly lower
than the amplitude it should have ($\sigma_B$).  Assume the Planck map
includes both modes labeled 1 and 2 and therefore has the expected
amplitude $\sigma_{353}$.  The amplitude of the gravity wave portion
of the BICEP2 map is now $\sigma_{GW} (\sigma_1/ \sigma_B)$, which
correlates only with the (1) modes in the Planck Map which have an
amplitude of $\sigma_{GW} (\sigma_1/ \sigma_B)$, thus the product of
the gravity wave modes in the two maps is: $\sigma_{GW}^2
(\sigma_1/ \sigma_B)^2$.  We have similar terms for gravitational
lensing modes and dust modes.  In equation (8) the numerator on the
right hand side is multiplied by a factor of $(\sigma_1/ \sigma_B)^2$
while the denominator is multiplied by a factor of
$(\sigma_1/ \sigma_B)$.  This multiplies $C_{PB}$ by a factor of
$(\sigma_1/ \sigma_B)$.  In equation 10 the left side of the
equation $C_{PB}(\sigma_{353}/\sigma_{B})$ is unchanged, because both
$C_{PB}$ and $\sigma_B$ have been multiplied by a factor of
$(\sigma_1/ \sigma_B)$ while $\sigma_{353}$ remains as before; this
leaves $x + z + 21.3y$ unchanged.  Our solutions for $x$, $z$, and $y$
remain unchanged as does our result for $r$.  Thus, our results are
not affected if the BICEP2 map drops some some modes.

The Planck team and the BICEP2 team have agreed to a joint analysis of
their data.  The BICEP2 team can ``observe'' the Planck 353 map using
their procedures which would involve dropping just the (2) modes.
Thus, their final ``Planck map'' would have an amplitude of
$\sigma_{353}^\prime = \sigma_{353} (\sigma_1/\sigma)$ because the
BICEP2 analyzed Planck map contains only the (1) modes.  The BICEP2
team would then find a correlation coefficient between their current
map with amplitude $\sigma_1$ and the new ``Planck Map'' with
amplitude of $\sigma_{353}^\prime$ of $C_{PB}^\prime
= \left(x\sigma_1^2 + z\sigma_1^2 + 21.3y\sigma_1^2\right)/
(\sigma_1\sigma_{353}^\prime)$, giving
$C_{PB}^\prime \sigma{353}^\prime/\sigma_1 = x + z + 21.3y$.  The
BICEP2 team will observe a correlation coefficient $C_{PB}^\prime$
with their new reduced ``Planck Map'' that is higher than we observe
by a factor of $\sigma_B/\sigma_1$, but this is compensated for
exactly by the fact that $\sigma_{353}^\prime$ is smaller than what we
observe by the same factor, so that they should get the same value of
$x + z + 21.3y$ that we get.  In the same way, we showed in our
comparison with Map I, that addition of ringing modes from outside the
BICEP2 region to the Planck map, by raising the amplitude of
$\sigma_{353}$ while simultaneously lowering the correlation
coefficient by the same factor leads to a similar estimated value of
$r = 0.106$.  Planck (Planck Collaboration 2014b) has (since our
original preprint was published on arXiv) produced an apodized map of
the BICEP2 region which is similar in construction to our Map IV.
They have not published this map, but have used it to make power
spectrum amplitude estimates of the {\it B} and \Emodes\ (from $40 < l
< 120$).  The ratio of their {\it E} and \Bmode\ powers in their map
is equal to what we get from our digitized Map IV to within $1\sigma$.
We observe a correlation coefficient of $C_{PB} = 0.181$ between the
BICEP2 map and our Map IV at 353 GHz.  We obtained a value of $r =
0.099$ with Map IV.  The Planck team simultaneously, upped their best
estimate for the ratio of the dust signal amplitude ratio
$\sigma_{353D}/ \sigma_BD$ to 24.5.  This is higher than our adopted
value of 21.3 and this has the effect of slightly raising the value of
$r$.  Applying both Map IV and the value of 24.5 leads to a best
estimate of $r = 0.104$.  The errors would be similar to our earlier
estimates.  Rounding as we did before and keeping significant figures
would give us $r = 0.10 \pm 0.04$, compared with our original estimate
of $r = 0.11 \pm 0.04$.  The difference is insignificant within the
errors.  Our error budget already included errors for the factor of
21.3, and for the errors associated with picking different maps for
comparison.

The power spectrum of the lowest polarization 30\% of the sky in
Planck at 353 GHz seems to hit noise at $C_l \sim 2.2 \times 10^{-3}
\uK^2$ at $l \sim 260$, where one is looking at noise amplitudes of
$\Delta_{BB}^2 \sim l(l+1)C_l/4\pi \sim 11.5 \uK^2$.  This noise
estimate in the \Bmodes\ is empirically based on when measurements of
the \Bmodes\ by Planck start to show uncertainty.  This very rough
estimate from the Planck power spectrum suggests $\sigma_{353N} \sim
3.4\uK$, similar to $\sigma_{353}$, suggesting noise makes a
significant contribution to the Planck map.  The Planck team notes that
to look for a gravity wave signal at $r = 0.1$ will require
subtraction of the dust signal.  This means correlating the two maps.
The fact that we observe a significant but low correlation with BICEP2
allows us to quantify the contribution of the dust signal.  Assuming
$\sigma_{353}$ was entirely due to dust (with no noise) would lead to
an overestimate of the dust contribution to the BICEP2 map.  Assuming
that the excess \Bmodes\ in the BICEP2 were due entirely to dust (and
not at all to gravity waves) would have produced correlations larger
than we observed.  Our correlation study allows us to quantify the
dust contribution to the BICEP2 signal, and we find it to be slightly
less than the gravity wave signal.

In conclusion our independent analysis shows evidence supporting a
detection of a gravity wave signal with $r = 0.11 \pm 0.04(1\sigma)$.
The Planck team and the BICEP2 team have agreed to a joint analysis of
their data.  It will be interesting to see if they reach similar
conclusions.  We look forward to that paper.

\section*{Acknowledgements}
  
We thank Joel Zinn, Bruce Draine, Lyman Page, Matias Zaldarriaga and
David Spergel for helpful conversations.

\label{lastpage}

\end{document}